\documentclass[longauth]{config/aa} 

\usepackage{comment}
\usepackage{graphicx}
\usepackage{multirow}
\usepackage{textgreek}
\usepackage{xargs}
\usepackage[dvipsnames]{xcolor}
\usepackage{xspace}
\usepackage{txfonts}
\usepackage{hyperref}
\hypersetup{
    colorlinks=true,
    linkcolor=blue,
    citecolor=blue,
    filecolor=magenta,      
    urlcolor=blue,
    }
\usepackage{subcaption}

\graphicspath{{./}{figs/}}
\usepackage{etoolbox}
\makeatletter
\newcommand\sendemail[3]{
\edef\@tempa{mailto:#1?subject=#2 }%
\edef\@tempb{\expandafter\html@spaces\@tempa\@empty}%
\href{\@tempb}{#3}}

\catcode\%=11
\def\html@spaces#1 #2{#1
\catcode\%=14
\makeatother

\newcommand{\citationneeded}{\textcolor{ForestGreen}{$^{\rm citation\;needed}$}}
\let\oldtextsigma\textsigma
\renewcommand{\textsigma}{\oldtextsigma\xspace}
\let\oldtextalpha\textalpha
\renewcommand{\textalpha}{\oldtextalpha\xspace}
\let\oldAA\AA
\renewcommand{\AA}{\text{\oldAA}\xspace}
\let\oldtextdegree\textdegree
\renewcommand{\textdegree}{\oldtextdegree\xspace}

\newcommand{\orcid}[2]{\href{http://orcid.org/#2}{#1{\includegraphics[height=10pt]{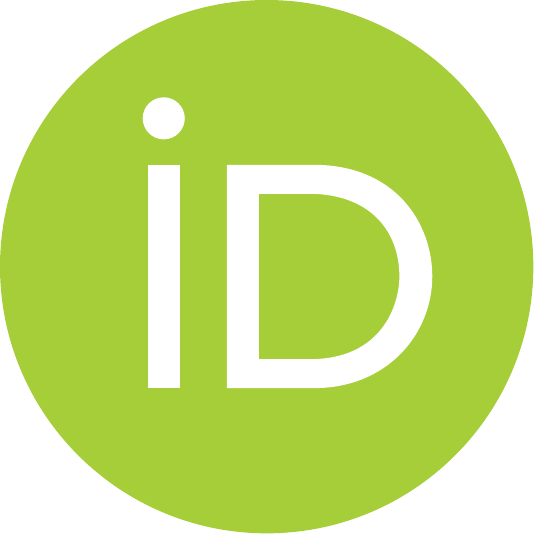}}}}

\newcommand{\target}{GS-z12\xspace}
\newcommand{\pox}{Pox~186\xspace}

\newcommand{\zspec}{12.48\xspace}

\newcommand{\kms}{\ensuremath{\mathrm{km\,s^{-1}}}\xspace}
\newcommand{\MSun}{\ensuremath{{\rm M}_\odot}\xspace}
\newcommand{\yr}{\ensuremath{{\rm yr}}\xspace}
\newcommand{\Myr}{\ensuremath{{\rm Myr}}\xspace}
\newcommand{\Gyr}{\ensuremath{{\rm Gyr}}\xspace}
\newcommand{\peryr}{\ensuremath{{\rm yr^{-1}}}\xspace}
\newcommand{\Lsun}{\hbox{\,${\rm L}_\odot$}}
\newcommand{\mum}{\text{\textmu m}\xspace}
\newcommand{\dex}{\text{dex}\xspace}
\newcommand{\kpc}{\text{kpc}\xspace}
\newcommand{\ZH}{\text{[Z/H]}\xspace}
\newcommand{\CO}{\text{[C/O]}\xspace}
\newcommand{\FeH}{\text{[Fe/H]}\xspace}
\newcommand{\percm}[1]{\ensuremath{\rm cm^{#1}}\xspace}

\newcommand{\eps}{\ensuremath{\epsilon}\xspace}
\newcommand{\mstar}{\ensuremath{M_\star}\xspace}
\newcommand{\mgas}{\ensuremath{M_\mathrm{gas}}\xspace}
\newcommand{\re}{\ensuremath{R_\mathrm{e}}\xspace}
\newcommand{\NHI}{\ensuremath{N_\mathrm{H\textsc{i}}}\xspace}
\newcommand{\tauv}{\ensuremath{\tau_\mathrm{V}}\xspace}
\newcommand{\beagletauv}{\ensuremath{\hat{\tau}_\mathrm{V}}\xspace}
\newcommand{\AV}{\ensuremath{A_\mathrm{V}}\xspace}
\newcommand{\xid}{\ensuremath{\xi_\mathrm{d}}\xspace}
\newcommand{\logoh}{\ensuremath{12 + \log\,(\mathrm{O/H})}\xspace}

\newcommand{\nelec}{\ensuremath{n_\mathrm{e}}\xspace}
\newcommandx{\Mout}[2][1=,2=]{\ensuremath{M_{\mathrm{out}{#2}}^{#1}}\xspace}
\newcommandx{\Mdotout}[2][1=,2=]{\ensuremath{\dot{M}_{\mathrm{out}{#2}}^{#1}}\xspace}

\newcommandx{\fluxdcgs}[1][1=-20]{\ensuremath{\mathrm{10^{#1}~erg~s^{-1}~cm^{-2}~\AA^{-1}}}\xspace}
\newcommandx{\fluxcgs}[1][1=-20]{\ensuremath{\mathrm{10^{#1}~erg~s^{-1}~cm^{-2}}}\xspace}
\newcommandx{\powercgs}[1][1=44]{$\times 10^{#1}$~erg~s$^{-1}$\xspace}
\newcommand{\Av}{\ensuremath{A_V}\xspace}

\newcommand{\jwst}{\textit{JWST}\xspace}
\newcommand{\hst}{\textit{HST}\xspace}
\newcommand{\ppxf}{{\sc ppxf}\xspace}
\newcommand{\beagle}{{\sc beagle}\xspace}
\newcommand{\forcepho}{{\sc forcepho}\xspace}
\newcommand{\prospector}{{\sc prospector}\xspace}

\defcitealias{curtis-lake+2023}{CLC23}
\defcitealias{nakajima+maiolino2022}{NM22}
\defcitealias{gutkin+2016}{G16}
\defcitealias{feltre+2016}{F16}

\newcommand{\Lyalpha}{\text{Ly\textalpha}\xspace}
\newcommand{\Halpha}{\text{H\textalpha}\xspace}
\newcommand{\Hbeta}{\text{H\textbeta}\xspace}
\newcommand{\Hgamma}{\text{H\textgamma}\xspace}
\newcommand{\Hdelta}{\text{H\textdelta}\xspace}
\newcommand{\Pabeta}{\text{Pa\textbeta}\xspace}
\newcommand{\Hepsilon}{\text{H\textepsilon}\xspace}

\newcommandx{\permittedEL}[6][1=O,2=III,3=,4=,5=,6=]{\text{{#1}\,{\sc {#2}}{#3}{#4}{#5}{#6}}\xspace}
\newcommandx{\semiforbiddenEL}[6][1=O,2=III,3=,4=,5=,6=]{\text{{#1}\,{\sc{#2}}]{#3}{#4}{#5}{#6}}\xspace}
\newcommandx{\forbiddenEL}[6][1=O,2=III,3=,4=,5=,6=]{\text{[{#1}\,{\sc{#2}}]{#3}{#4}{#5}{#6}}\xspace}

\newcommand{\EW}[1]{\text{EW(#1)}\xspace}

\newcommand{\HI}{\permittedEL[H][i]}
\newcommand{\HII}{\permittedEL[H][ii]}

\newcommand{\NV}{\permittedEL[N][v]}
\newcommandx{\NVL}[1][1=1243]{\permittedEL[N][v][\textlambda][#1]}
\newcommandx{\NVall}{\permittedEL[N][v][\textlambda][\textlambda][1239,][1243]}

\newcommandx{\CIIL}[1][1=1334]{\permittedEL[C][ii][\textlambda][#1]}

\newcommand{\NIV}{\semiforbiddenEL[N][iv]}
\newcommandx{\NIVL}[1][1=1486]{\semiforbiddenEL[N][iv][\textlambda][#1]}

\newcommand{\CIV}{\permittedEL[C][iv]}
\newcommandx{\CIVL}[1][1=1550]{\permittedEL[C][iv][\textlambda][#1]}
\newcommand{\CIVall}{\permittedEL[C][iv][\textlambda][\textlambda][1549,][1551]}

\newcommand{\HeII}{\permittedEL[He][ii]}
\newcommandx{\HeIIL}[1][1=1640]{\permittedEL[He][ii][\textlambda][#1]}

\newcommand{\OIII}{\semiforbiddenEL[O][iii]}
\newcommandx{\OIIIL}[1][1=1666]{\semiforbiddenEL[O][iii][\textlambda][#1]}
\newcommand{\OIIIall}{\semiforbiddenEL[O][iii][\textlambda][\textlambda][1661,][1666]}

\newcommand{\NIII}{\semiforbiddenEL[N][iii]}
\newcommandx{\NIIIL}[1][1=1750]{\semiforbiddenEL[N][iii][\textlambda][#1]}
\newcommand{\NIIIall}{\semiforbiddenEL[N][iii][\textlambda][\textlambda][1747--][1754]}

\newcommandx{\CIII}{\semiforbiddenEL[C][iii]}
\newcommandx{\CIIIL}[1][1=1909]{\semiforbiddenEL[C][iii][\textlambda][#1]}
\newcommand{\CIIIall}{\semiforbiddenEL[C][iii][\textlambda][\textlambda][1907,][1909]}

\newcommand{\NeIV}{\forbiddenEL[Ne][iv]}
\newcommandx{\NeIVL}[1][1=2424]{\forbiddenEL[Ne][iv][\textlambda][#1]}
\newcommand{\NeIVall}{\forbiddenEL[Ne][iv][\textlambda][\textlambda][2422,][2424]}

\newcommand{\MgII}{\permittedEL[Mg][ii]}
\newcommandx{\MgIIL}[1][1=2803]{\permittedEL[Mg][ii][\textlambda][#1]}
\newcommand{\MgIIall}{\permittedEL[Mg][ii][\textlambda][\textlambda][2796,][2803]}

\newcommand{\NeV}{\forbiddenEL[Ne][v]}
\newcommandx{\NeVL}[1][1=3426]{\forbiddenEL[Ne][v][\textlambda][#1]}
\newcommand{\NeVall}{\forbiddenEL[Ne][v][\textlambda][\textlambda][3346,][3426]}

\newcommand{\OII}{\forbiddenEL[O][ii]}
\newcommandx{\OIIL}[1][1=3727]{\forbiddenEL[O][ii][\textlambda][#1]}
\newcommand{\OIIall}{\forbiddenEL[O][ii][\textlambda][\textlambda][3726,][3729]}

\newcommand{\NeIII}{\forbiddenEL[Ne][iii]}
\newcommandx{\NeIIIL}[1][1=3869]{\forbiddenEL[Ne][iii][\textlambda][#1]}
\newcommand{\NeIIIall}{\forbiddenEL[Ne][iii][\textlambda][\textlambda][3869,][39xx]}

\newcommand{\hda}{\ensuremath{\mathrm{H\text{\textdelta}_A}}\xspace}
\newcommand{\hga}{\ensuremath{\mathrm{H\text{\textgamma}_A}}\xspace}

\begin{document} 

   \title{JADES: Carbon enrichment 350~Myr after the Big Bang in a gas-rich galaxy}

   \titlerunning{High C/O in \target}

\author{
\orcid{Francesco D'Eugenio}{0000-0003-2388-8172}
\inst{\hyperlink{aff1}{1}}\fnmsep\inst{\hyperlink{aff2}{2}}\thanks{E-mail: francesco.deugenio@gmail.com} 
\and
\orcid{Roberto Maiolino}{0000-0002-4985-3819}
\inst{\hyperlink{aff1}{1}}\fnmsep\inst{\hyperlink{aff2}{2}}\fnmsep\inst{\hyperlink{aff3}{3}}
\and
\orcid{Stefano Carniani}{0000-0002-6719-380X}
\inst{\hyperlink{aff4}{4}}
\and
\orcid{Emma Curtis-Lake}{0000-0002-9551-0534}
\inst{\hyperlink{aff5}{5}}
\and
\orcid{Joris Witstok}{}
\inst{\hyperlink{aff1}{1}}\fnmsep\inst{\hyperlink{aff2}{2}}
\and
\orcid{Jacopo Chevallard}{0000-0002-7636-0534}
\inst{\hyperlink{aff6}{6}}
\and
\orcid{Stephane Charlot}{0000-0003-3458-2275}
\inst{\hyperlink{aff7}{7}}
\and
\orcid{William M. Baker}{0000-0003-0215-1104}
\inst{\hyperlink{aff1}{1}}\fnmsep\inst{\hyperlink{aff2}{2}}
\and
\orcid{Santiago Arribas}{0000-0001-7997-1640}
\inst{\hyperlink{aff8}{8}}
\and
\orcid{Kristan Boyett}{0000-0003-4109-304X}
\inst{\hyperlink{aff9}{9}}\fnmsep\inst{\hyperlink{aff10}{10}}
\and
\orcid{Andrew J.\ Bunker }{0000-0002-8651-9879}
\inst{\hyperlink{aff6}{6}}
\and
\orcid{Mirko Curti}{0000-0002-2678-2560}
\inst{\hyperlink{aff11}{11}}\fnmsep\inst{\hyperlink{aff1}{1}}\fnmsep\inst{\hyperlink{aff2}{2}}
\and
\orcid{Daniel J.\ Eisenstein}{0000-0002-2929-3121}
\inst{\hyperlink{aff12}{12}}
\and
\orcid{Kevin Hainline}{0000-0003-4565-8239}
\inst{\hyperlink{aff13}{13}}
\and
\orcid{Zhiyuan Ji}{0000-0001-7673-2257}
\inst{\hyperlink{aff13}{13}}
\and
\orcid{Benjamin D.\ Johnson}{0000-0002-9280-7594}
\inst{\hyperlink{aff12}{12}}
\and
\orcid{Tobias J. Looser}{0000-0002-3642-2446}
\inst{\hyperlink{aff1}{1}}\fnmsep\inst{\hyperlink{aff2}{2}}
\and
\orcid{Kimihiko Nakajima}{0000-0003-2965-5070}
\inst{\hyperlink{aff14}{14}}
\and
\orcid{Erica Nelson}{0000-0002-7524-374X}
\inst{\hyperlink{aff15}{15}}
\and
\orcid{Marcia Rieke}{0000-0002-7893-6170}
\inst{\hyperlink{aff13}{13}}
\and
\orcid{Brant Robertson}{0000-0002-4271-0364}
\inst{\hyperlink{aff16}{16}}
\and
\orcid{Jan Scholtz}{}
\inst{\hyperlink{aff1}{1}}\fnmsep\inst{\hyperlink{aff2}{2}}
\and
\orcid{Renske Smit}{0000-0001-8034-7802}
\inst{\hyperlink{aff17}{17}}
\and
\orcid{Giacomo Venturi}{0000-0001-8349-3055}
\inst{\hyperlink{aff4}{4}}
\and
\orcid{Sandro Tacchella}{0000-0002-8224-4505}
\inst{\hyperlink{aff1}{1}}\fnmsep\inst{\hyperlink{aff2}{2}}
\and
\orcid{Hannah \"Ubler}{0000-0003-4891-0794}
\inst{\hyperlink{aff1}{1}}\fnmsep\inst{\hyperlink{aff2}{2}}
\and
\orcid{Christopher N. A. Willmer}{0000-0001-9262-9997}
\inst{\hyperlink{aff13}{13}}
\and
\orcid{Chris Willott}{0000-0002-4201-7367}
\inst{\hyperlink{aff18}{18}}
}

\institute{
\hypertarget{aff1}{Kavli Institute for Cosmology, University of Cambridge, Madingley Road, Cambridge, CB3 0HA, UK}\\
\and
\hypertarget{aff2}{Cavendish Laboratory, University of Cambridge, 19 JJ Thomson Avenue, Cambridge, CB3 0HE, UK}\\
\and
\hypertarget{aff3}{Department of Physics and Astronomy, University College London, Gower Street, London WC1E 6BT, UK}\\
\and
\hypertarget{aff4}{Scuola Normale Superiore, Piazza dei Cavalieri 7, I-56126 Pisa, Italy}\\
\and
\hypertarget{aff5}{Centre for Astrophysics Research, Department of Physics, Astronomy and Mathematics, University of Hertfordshire, Hatfield AL10 9AB, UK}\\
\and
\hypertarget{aff6}{Department of Physics, University of Oxford, Denys Wilkinson Building, Keble Road, Oxford OX1 3RH, UK}\\
\and
\hypertarget{aff7}{Sorbonne Universit\'e, CNRS, UMR 7095, Institut d'Astrophysique de Paris, 98 bis bd Arago, 75014 Paris, France}\\
\and
\hypertarget{aff8}{Centro de Astrobiolog\'ia (CAB), CSIC–INTA, Cra. de Ajalvir Km.~4, 28850- Torrej\'on de Ardoz, Madrid, Spain}\\
\and
\hypertarget{aff9}{School of Physics, University of Melbourne, Parkville 3010, VIC, Australia}
\and
\hypertarget{aff10}{ARC Centre of Excellence for All Sky Astrophysics in 3 Dimensions (ASTRO 3D), Australia}
\and
\hypertarget{aff11}{European Southern Observatory, Karl-Schwarzschild-Strasse 2, 85748 Garching, Germany}\\
\and
\hypertarget{aff12}{Center for Astrophysics $|$ Harvard \& Smithsonian, 60 Garden St., Cambridge MA 02138 USA}\\
\and
\hypertarget{aff13}{Steward Observatory, University of Arizona, 933 North Cherry Avenue, Tucson, AZ 85721, USA}\\
\and
\hypertarget{aff14}{National Astronomical Observatory of Japan, 2-21-1 Osawa, Mitaka, Tokyo 181-8588, Japan}\\
\and
\hypertarget{aff15}{Department for Astrophysical and Planetary Science, University of Colorado, Boulder, CO 80309, USA}
\and
\hypertarget{aff16}{Department of Astronomy and Astrophysics, University of California, Santa Cruz, 1156 High Street, Santa Cruz, CA 95064, USA}\\
\and
\hypertarget{aff17}{Astrophysics Research Institute, Liverpool John Moores University, 146 Brownlow Hill, Liverpool L3 5RF, UK}
\and
\hypertarget{aff18}{NRC Herzberg, 5071 West Saanich Rd, Victoria, BC V9E 2E7, Canada}\\
}

   \authorrunning{F. D'Eugenio et al.}
   \date{}

  \abstract{
Finding the emergence of the first generation of metals in the early Universe, and identifying their origin, are some of the most important goals of modern astrophysics. 
We present deep \jwst/NIRSpec spectroscopy of \target, a galaxy at z=12.5, in which  we report the detection of \CIIIall 
   nebular emission.
   This is the most distant detection of a metal transition and the most distant redshift determination via  emission lines.
   In addition, we report tentative detections of \OIIall and \NeIIIL, and possibly \OIIIall.
   By using the accurate redshift from \CIII, we can  model the \Lyalpha drop to 
   reliably measure an absorbing column density of hydrogen of $\NHI\approx 10^{22}~\percm{-2}$ --
   too high for an IGM origin and implying abundant ISM in \target or CGM around it. We infer a lower limit for the neutral gas mass of about $10^7~M_\odot$ which, compared with a stellar mass of $\sim 4\times 10^7~M_\odot$ inferred from the continuum fitting, implies a gas fraction higher than about 0.1--0.5.
   We derive a 
   solar or even super-solar carbon-to-oxygen ratio, tentatively $\CO>0.15$. This is higher than the C/O measured in galaxies discovered by \jwst at $z=6\text{--}9$,
    and higher
   than the C/O arising from 
   Type-II supernovae enrichment, while AGB stars cannot contribute to carbon enrichment at these early epochs and low metallicities. Such a high C/O in a galaxy
   observed 350~Myr after the Big Bang may be explained by the yields of
   extremely metal poor stars, and may even be the heritage of the
   first generation of supernovae from Population III progenitors.
   }


   \maketitle
%

\section{Introduction}

The appearance of the first galaxies marks a key phase transition 
of the Universe, i.e. the end of the dark ages.
A keystone of this phase transition is the start of
stellar nucleosynthesis and the diffusion of metals.
Extensive theoretical work has been devoted to
predicting the properties of the first generation of stars
\citep[Population~III, hereafter: PopIII; e.g.,][]{hirano+2014}
and their supernova yields \citep{heger+woosley2010,limongi+chieffi2018}.
While PopIII stars are thought to be short-lived \citep[with 
typical masses of 10--40~\MSun;][]{hosokawa+2011},
the chemical `signature' of their yields may be still observable today.
Empirically, extensive searches of PopIII-enriched systems have 
been undertaken both within the Milky Way
\citep[e.g.,][]{frebel+norris2015}
and in extragalactic absorbers, including both the most
metal-poor damped \Lyalpha (DLA) systems \citep[e.g.,][]{pettini+2008,salvadori+ferrara2012,cooke+2017,saccardi+2023}
and Lyman-limit systems \citep[LLS;][]{fumagalli+2016,saccardi+2023}.

The launch of \jwst enabled, for the first time, the measurement of the physical
properties of galaxies out to z$>$10 \citep{curtis-lake+2023, robertson+2023, bunker+2023a,tacchella+2023,maiolino+2023,arrabal-haro+2023,hsiao+2023}.
These high-redshift 
observations challenge the extrapolation of trends derived at lower redshifts. These are generally understood
in terms of decreasing gas metallicity
\citep{schaerer+2023, curti+2023b, nakajima+2023}
and increasing density \citep{reddy+2023},
ionisation parameter \citep{cameron+2023b},
temperature \citep{curti+2023a} and stochasticity of their star-formation histories \citep[SFH;][]{dressler+2023, endsley+2023, looser+2023}.
However, extrapolations
of these trends at z$>$10 do not fully explain the observed properties of
galaxies. One reason is that as our observations approach the end of 
`Cosmic Dawn' \citep[150--250~Myr after the Big Bang,][]{robertson2022},
galaxies may carry stronger imprints from the
first generation of metal-poor stars, whose physical properties and spectra are poorly understood.
Most notably, nebular emission-line ratios in a
luminous galaxy at $z=10.6$
\citep[GN-z11,][]{oesch+2016, bunker+2023a}
seem to require exotic chemical abundances \citep{cameron+2023a},
or proto-globular clusters \citep{senchyna+2023},
or Wolf-Rayet stars -- possibly
with a fine-tuned SFH \citep{kobayashi+ferrara2023}. Moreover, a supermassive accreting black hole has been identified in this galaxy, suggesting that the peculiar chemical abundances might be primarily associated with its nuclear region \citep{maiolino+2023}.

In addition to rare objects like the remarkably 
bright GN-z11, \jwst enabled the observations of spectra of
regular galaxies at $z=10\text{--}13$ \citetext{\citealp{curtis-lake+2023},
hereafter: \citetalias{curtis-lake+2023}; \citealp{hsiao+2023};
\citealp{arrabal-haro+2023}}, at redshifts even higher than GN-z11, i.e.,
even nearer to the end of Cosmic Dawn.
While less extreme than objects like GN-z11, 
these galaxies are more representative of the typical 
physical properties of galaxies at those epochs
\citetext{\citetalias{curtis-lake+2023},
\citealp{robertson+2023}}.
One of the remarkable findings of
\citetalias{curtis-lake+2023} was
the lack of any emission lines -- even with the unprecedented
depth of the \jwst Advanced Deep Extragalactic Survey
\citep[JADES;][]{eisenstein+2023a, rieke+2023, bunker+2023b}.
Unfortunately, the lack of emission lines severely affects our ability to constrain
the physical properties and chemical abundances of these galaxies, because systematic uncertainties dominate the constraints \citetext{introduced
through strong priors on, e.g., the shape of the
SFH as well as other factors; \citetalias{curtis-lake+2023}}.

To address this shortcoming, the large programme PID~3215
\citep{eisenstein+2023b}, while obtaining deep multi-band imaging, has in parallel also obtained the deepest spectroscopic 
observations yet of galaxies at $z>10$, thanks to a 50-hours integration with \jwst/NIRSpec, for a 
5-\textsigma emission-line sensitivity of $10\text{--}5\times \fluxcgs[-20]$ at $\approx 2\text{--}5~\mum$ (for spectrally unresolved lines).
In this article, we report the first analysis of the new
spectroscopic data for \target, a $z>10$ galaxy already 
analysed in \citetalias{curtis-lake+2023} and 
\citet{robertson+2023}. After
presenting the data reduction and analysis in \S~\ref{s.dran},
we show the physical constraints obtained from the data
(\S~\ref{s.stellar}-\ref{s.diagn}) and conclude with a brief discussion and
outlook (\S~\ref{s.discon}).

Throughout this work, we assume the \citet{planck+2020} 
cosmology, a \citet{chabrier2003} initial mass function (IMF) with an upper-mass cutoff of
300~\MSun,
and the Solar abundances of \citet{asplund+2009}.
Stellar masses refer to the total stellar mass formed (i.e., the integral of a galaxy SFH), and distances are proper distances.

\section{Observations, sample and data analysis}\label{s.dran}

\begin{figure*}
   \includegraphics[width=\textwidth]{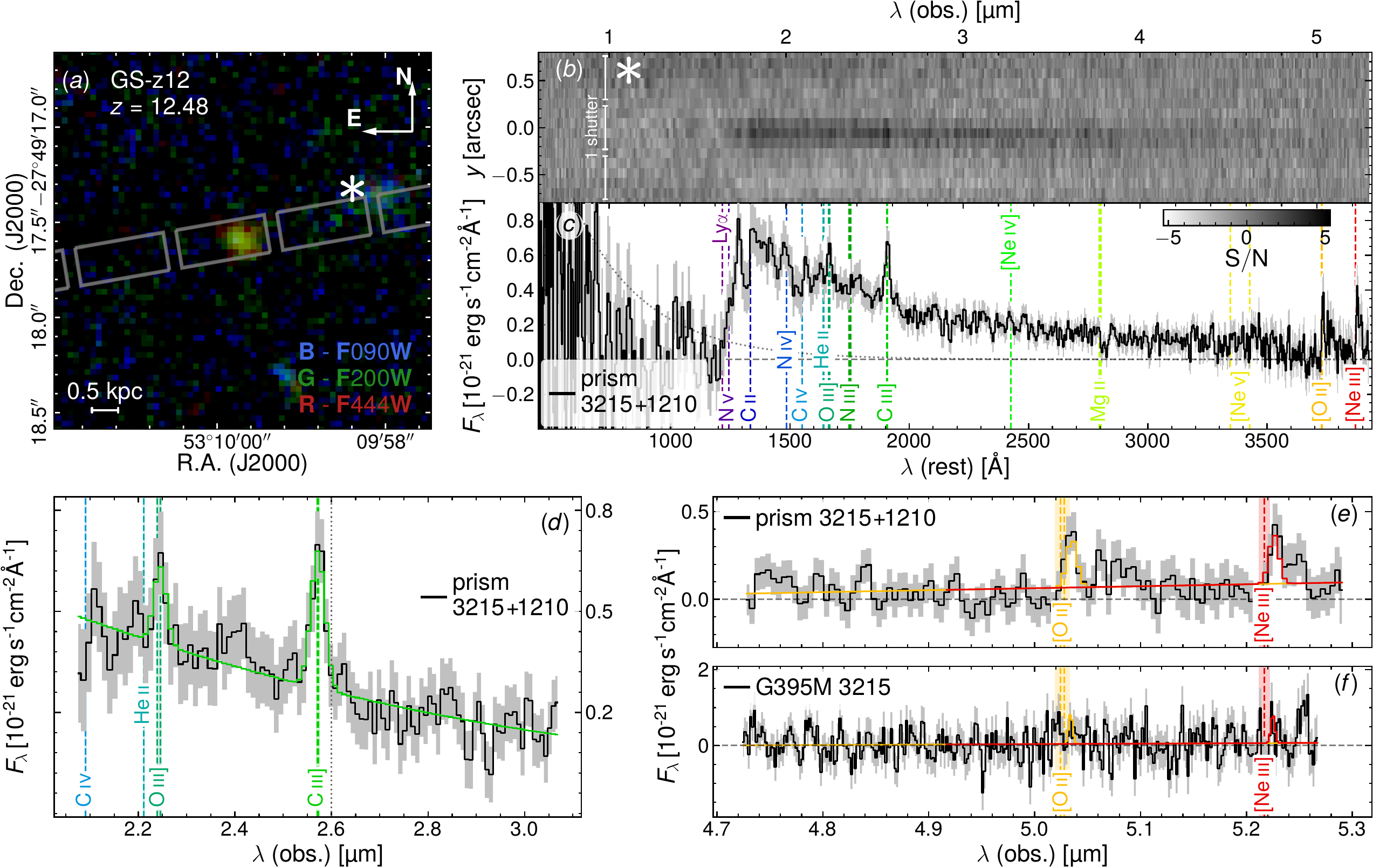}
   {\phantomsubcaption\label{f.data.a}
    \phantomsubcaption\label{f.data.b}
    \phantomsubcaption\label{f.data.c}
    \phantomsubcaption\label{f.data.d}
    \phantomsubcaption\label{f.data.e}
    \phantomsubcaption\label{f.data.f}
   }
   \caption{Panel~\subref{f.data.a}; false-colour RGB image, highlighting
   the position of the NIRSpec/MSA shutters in PID~3215. The
   asterisk indicates the position of an interloper with low
   surface brightness, which was removed in the data reduction.
   Panel~\subref{f.data.b}; 2-d S/N map, showing the three
   central shutters (the asterisk is the position of the interloper).
   Panel~\subref{f.data.c}; 1-d boxcar-extracted prism spectrum, combining 3215 and 1210;
   the vertical dashed lines marking the position of
   strong emission lines at $z=\zspec$.
   Panels~\subref{f.data.d}--\subref{f.data.f};
   combined 3215 and 1210 data and model spectrum around the
   \OIII, \CIII, \OII and \NeIII lines, for the prism
   (panels~\subref{f.data.d} and~\subref{f.data.e})
   and for the G395M grating (only 3215; panel~\subref{f.data.f}).
   \CIII is detected at the 5-\textsigma level (7-\textsigma with the bootstrapping method).
   \OIII is not robustly detected (2.3-\textsigma significance); \OII and \NeIII are marginally detected only in 
   the prism (4- and 3.5-\textsigma) but not in the grating,
   despite comparable sensitivity. The vertical dashed lines mark the
   wavelength of the emission lines at the redshift of the object, with
   the shaded region indicating the redshift uncertainty.
   }
   \label{f.data}
\end{figure*}

The observations consist of NIRSpec Micro-Shutter Assembly (MSA) spectroscopy
with the prism and with the G140M and G395M gratings. These data were obtained as part of programmes PID~1210 \citetext{PI~N.~L\"utzgendorf; already presented in
\citetalias{curtis-lake+2023} and \citealp{bunker+2023b}} and PID~3215
\citetext{PI~D.~Eisenstein and R.~Maiolino, \citealp{eisenstein+2023b}}.
A summary of the observing configurations and total integrations
is provided in Table~\ref{t.obs}.
The data reduction was performed exactly as described in \citet{bunker+2023b, carniani+2023}.
We used nodding for background subtraction, and extracted the
1-d spectrum using a 3-pixel window.
Effective line spread functions (LSF) were obtained
from modelling the instrument, as described in \citet{degraaff+2023}. The input
model for each galaxy was obtained from \textsc{forcepho} (Johnson et~al., in~prep.),
using the same methods as described in \citet{baker+2023}. The results are
shown in Table~\ref{t.physpar}.
We applied a slit-loss 
correction
appropriate for point sources (cf. the morphological parameters reported in Table~\ref{t.physpar}). We note that this slit-loss 
correction is optimised for a 5-pixel aperture.
For the analysis, we combine the data from PIDs~3215 and~1210 using a simple
inverse-variance weighting.

\begin{table}
  \centering
  Summary of the observations from programmes 1210 and 3215.
  \begin{tabular}{lccc}
  \hline
  Disperser               & prism      & G140M & G395M \\
  Filter                  & CLEAR      & F070LP& F290LP\\
  Spectral resolution $R$ & 30--300    & 700--1500    & 700--1500           \\
  Exp. time 1210  [h]     & 18.7       & 4.7          & 4.7                 \\
  Exp. time 3215  [h]     & 46.7       & 11.7         & 46.7 (37.4)$^\ddag$ \\
  Exp. time Total [h]     & 65.4       & 16.4         & 51.5 (42.1)$^\ddag$ \\
  \hline
  \end{tabular}
  \caption{
  These exposure times apply only to \target; other
  targets in the two programmes may have different 
  exposure times, depending on the mask allocation.
  $^\ddag$ The last visit was affected by short circuits, and the true exposure time in
  PID~3215 is reduced to 37.4~h.
  }\label{t.obs}
\end{table}

In the left column of Fig.~\ref{f.data} we display NIRCam imaging
for \target (panel~\subref{f.data.a}),
with overlaid the configuration of the NIRSpec MSA
shutters in PID~3215 (these observations consist of five dithered pointings).
The 2-d signal-to-noise ratio (S/N) and the boxcar extracted 1-d spectrum
are shown in panels~\subref{f.data.b} and~\subref{f.data.c}.

Panel~\subref{f.data.a} shows a contaminant with
low surface brightness to the west (marked by
an asterisk). This contaminant also appears at the
top of the 2-d S/N map (panel~\ref{f.data.b}), and its presence was taken into account in the 
data reduction.
Nevertheless, for the prism data of 3215, we measured a negative flux blueward of the \Lyalpha drop. The origin of this artefact is still
unclear.
To remove it, we extracted two background
spectra from the 2-d spectrum (above and below the object trace),
fitted a declining
exponential to these data, and removed the average best-fit
background model from the 1-d spectrum.
This procedure has no detectable impact on the 
emission-line fluxes, but affects the shape of
the continuum and \Lyalpha drop.
The uncertainties from the data reduction pipeline
were validated by inspecting the dispersion of the
individual integrations in each wavelength channel
(Appendix~\ref{a.data}). Overall, we find that the data-reduction
pipeline gives more conservative noise estimates than bootstrapping,
which is expected because bootstrapping does not take into account
correlated noise.

\subsection{Emission-line fitting}\label{s.elf}

Emission-line fluxes and equivalent widths (EW) were measured using a local pixel-integrated 
Gaussian model with
linear background, in a window of 0.3~\mum on either
side of the expected line location. The lines were 
assumed to be spectrally unresolved (Gaussian 
\textsigma equal to one spectral pixel). The model
has four free parameters: flux, redshift, and two
coefficients for the background. The redshift
was constrained to be centred at $z=12.479$ using a 
Gaussian prior with dispersion set to 0.014
(this redshift dispersion is the uncertainty we derived from
the redshift estimate, see \S~\ref{s.c3}). The fiducial values and
uncertainties were estimated using a
Markov-Chain Monte-Carlo integrator
\citep{foreman-mackey+2013}; the results are 
reported in Table~\ref{t.emlines}.

The~four~emission~lines~\OIIIall,~\CIIIall, 
\OIIall and \NeIIIL (hereafter simply: \OIII, 
\CIII, \OII and \NeIII) are central to our analysis of chemical
abundances, and for this reason are measured adopting a different strategy 
compared to all the other lines. In addition, while \OIII and \CIII 
are observed only by the prism, \OII and \NeIII are covered both by the prism and by the G395M grating. Therefore, to take full advantage of the
100-h total integration for \OII and \NeIII, we fit simultaneously the
prism and grating spectra.
The model uses six pixel-integrated Gaussians, one each for \OIII and \CIII 
(only present in the prism spectrum), and two each for \OII and \NeIII 
(which require separate models for the prism and grating spectra).
In total, the model has sixteen free parameters.
\CIII has three free parameters (flux, width, and redshift for the prism 
only); \OIII has two free parameters, flux and redshift for the prism only;
its width is constrained to be the same as \CIII,
and its redshift has a Gaussian prior centred on the \CIII redshift and with $\sigma_z = 0.02$.
\OII and \NeIII are treated as unresolved, hence have a fixed Gaussian
dispersion equal to one spectral pixel (as appropriate for the prism and 
grating separately). 
The redshifts of \OII are free and independent in the prism and grating, but we 
apply Gaussian priors of width $\sigma_z = 0.02$ to maintain them consistent 
with \CIII. This gives two free parameters (with strong priors). The redshifts 
of \NeIII are the same as for \OII in each of the two disperser configurations.
For both \OII and \NeIII, the prism flux is a free parameter, while the grating flux
is set to the prism flux up-scaled by 11 per~cent \citep[to take into account the systematic flux calibration discrepancy between these two observing modes, e.g.,][]{bunker+2023a}. In total,
we have nine free parameters describing six Gaussian line shapes. We further add seven parameters
to model the local background. The background around \CIII and 
\OIII in the prism is modelled with a 2\textsuperscript{nd}-order polynomial (three free parameters); the background around \OII and \NeIII is modelled
with a straight line, requiring two parameters each for the prism and for the
grating spectrum.
The prism spectral resolution and detector 
sampling prevent us from resolving the two
variable-ratio doublets \CIII and \OII.
We therefore adopt wavelengths of 
1,907.71 and 3,728.49~\AA, respectively, obtained by averaging the vacuum
wavelengths of the two lines in each doublet. Depending on the doublet ratio
for \CIII and \OII, this choice introduces
a systematic error of up to 0.1 and
0.4 pixels.

A summary of the emission lines and of their fluxes (or upper limits) is provided in Table~\ref{t.emlines}.

\subsection{Detection of \texorpdfstring{\CIII}{CIII}}\label{s.c3}

We adopt a detection threshold of 5~\textsigma, and report the detection of an emission line 
at $2.57~\mum$, with a 5-\textsigma significance. This S/N estimate is based on the
conservative uncertainties reported by the data reduction pipeline (bootstrapping the data
to estimate the uncertanties would increase the significance to 7~\textsigma,
as discussed in Appendix~\ref{a.data}).
At the redshift initially estimated by \citetalias{curtis-lake+2023} from the \Lyalpha drop 
($z=12.6\pm0.05$),
the line emission at 2.57~\mum is significantly offset (2,600~\kms, i.e. one spectral resolution element) from the
expected location of \CIII (the closest emission line; the expected location
is highlighted by the vertical dotted line in Fig.~\ref{f.data.d}).
However, in addition to the formal significance, other pieces of evidence contribute to confirming that this is a solid detection.
The comparison of the two independent datasets 3215 and 1210 (Fig.~\ref{f.integs.g}) shows that the line 
is seen independently, at the expected level of significance, in both observations. The robustness of the line detection is also confirmed by the visual inspection of the 2-d spectrum, in which the line is observed on three pixels in the spatial direction along the slit.  Based on these various lines of evidence, we interpret this line as \CIII, making it the most distant metal line detection to date.

We measure a high $\EW{\CIII}=30\pm7~\AA$.
Similarly high \EW{\CIII} have been found in some local metal-poor dwarf galaxies, such as \pox \citep[$\mstar\approx 10^5\,\MSun$;][]{kunth+1981,kumari+2023},
considered a local analogue of high-redshift star-forming galaxies, as well as in more massive galaxies at intermediate redshifts \citep[$2<z<4$;][]{lefevre+2019} and at
$7<z<9$ \citep{stark+2017}.

With our joint modelling fit, i.e. including the tentative detections of other lines discussed in the next section, we find a redshift of
$z = 12.482\pm0.012$. Modelling only \CIII at the mean rest-frame wavelength
of 1,907.71~\AA, we find $z=12.479\pm0.014$.
In both cases, there is a clear shift with
respect to the earlier redshift measurement based off
the observed wavelength of the \Lyalpha drop \citepalias{curtis-lake+2023}.
We interpret this redshift discrepancy
as evidence for a high 
column density of cold gas in the inter-stellar medium (ISM) of \target, or in its immediate surroundings, whose \Lyalpha damping wings affect the \Lyalpha drop; this will be further discussed in \S~\ref{s.coldgas}.

\subsection{Additional tentative detections and upper limits}\label{s.ne3o2}

In Table~\ref{t.emlines} we also report a tentative detection of \OIII
(2.3-\textsigma significance, or 3.5-\textsigma if adopting the bootstrap method discussed in the Appendix). This line is found at a wavelength consistent
with the redshift of \CIII, but it is not detected in the
1210 and 3215 observations taken separately (Fig.\ref{f.integs}).

Similarly, there appears to be a \CIVall (hereafter: \CIV) P-Cygni line at rest-frame 1,550~\AA, and a blue-shifted absorption of \CIIL. If confirmed, these might trace a galactic outflow, but the high implied velocities of about 3,000~\kms would require an AGN. On the basis of current data, these features are considered undetected.

Table~\ref{t.emlines} also reports tentative detections of \OII (at 4.4~\textsigma)
and \NeIII (at 3.7~\textsigma). Both lines are below the adopted threshold of 5~\textsigma. 
In addition to this threshold criterion, we do not consider these lines to be
secure detections for the reasons discussed in detail in Appendix~\ref{a.data}. The arguments in
favour of a detection are the correct
inter-line wavelength separation, and the
combined S/N from the two lines.
Yet, several arguments cast doubts about their detection:
(i) the moderate (2-\textsigma) tension 
between the redshifts of \CIII and
\OII--\NeIII (the dashed
vertical lines and shaded region in
Fig.~\ref{f.data.f} show the expected
location of these lines given the
\CIII redshift and its associated uncertainty). (ii) the unphysical wavelength offset of 2~pixels of \OII measured in 1210 with respect to 3215
(cf. cyan and sand lines in Fig.~\ref{f.integs.h}). (iii) while \NeIII appears in both datasets at the same wavelength,
its profile in 1210 appears to be too
narrow and is consistent with a noise spike.
(iv) visually inspecting the
2-d S/N map (Fig.~\ref{f.data.b}), we find no
strong evidence for either \OII or \NeIII (unlike
for \CIII, which is clearly visible); \OII, in particular,
appears to originate from a single spaxel, which
would be unphysical, because at 5~\mum the \jwst point
spread function is well sampled by the NIRSpec pixels, so we would expect the line to
be spatially extended (like \CIII).

The above difficulties with the 4- and 3-\textsigma tentative detections justify our choice
of a 5-\textsigma detection threshold. Based on this value, we only consider the 
\CIII detection as solid. We adopt 3-\textsigma upper limits (99.9~per cent confidence) on \CIV, \OII and \NeIII.
Nevertheless, promoting \OII and \NeIII to detections would not alter our conclusions. To help the reader judge the effect of assuming both \OII and \NeIII to be detected, where relevant we report the results when assuming detections, marking them with a small $\dagger$
symbol (Figs~\ref{f.coab} and~\ref{f.cooh}).

\begin{table}
    \centering

    Physical parameters of \target.\\
    \begin{tabular}{llcc}
    
    \hline
    \multirow{4}{*}{\rotatebox[origin=c]{90}{\forcepho}}
    & \re            & [arcsec]       & $0.040\pm0.003$ \\
    & P.A.           & [degree]       & $70\pm6$        \\
    & axis ratio $q$ & ---            & $0.55\pm0.07$   \\
    & S\'ersic index & ---            & $0.90\pm0.09$   \\
    \hline
    \multirow{4}{*}{\rotatebox[origin=c]{90}{\beagle}}
    & $\log\,\mstar$ & [\dex \MSun]   & $7.64\pm0.15$ \\
    & SFR            & [\MSun~\peryr] & $1.15\pm0.15$            \\
    & \beagletauv    & ---            & $0.12\pm0.04$          \\
    & \logoh         & [\dex]         & $7.9\pm0.2$            \\
    \hline
    \end{tabular}
    \caption{\forcepho photometry was obtained as explained in
    \citet{baker+2023}. The \beagle setup is described in \S~\ref{s.stellar}.
    }\label{t.physpar}
\end{table}
\begin{table}
    \centering
    \setlength{\tabcolsep}{4pt}
    Nebular emission lines for \target.
    \begin{tabular}{l|lcc}
  \hline
    & Line(s)            & Flux         & EW \\
  \hline
   \multirow{11}{*}{PRISM}
    & \NIVL               & $<3.3$                & $<6.4$                \\
    & \CIVall             & $<3.3$                & $<5.5$                \\
    & \HeIIL              & $<2.6$                & $<4.2$                \\
    & \OIIIall$^\ddag$    & $(6.0\pm2.6)$         & $8\pm3$              \\
    & \NIIIall            & $<2.2$                & $<3.8$                \\
    & \CIIIall$^\ddag$    & $\mathbf{12.4\pm2.5}$ & $\mathbf{30\pm7}$    \\
    & \NeIVall            & $<1.0$                & $<3.9$                \\
    & \MgIIall            & $<0.8$                & $<4.6$                \\
    & \NeVL               & $<1.0$                & $<8.6$                \\
    & \OIIall$^\ddag$     & $(3.5\pm0.8)$         & $41\pm12$            \\
    & \NeIIIL$^\ddag$     & $(3.3\pm0.9)$         & $39\pm13$            \\
  \hline
   \multirow{2}{*}{G140M}
    & \Lyalpha            & $<5.8$                & ---            \\
    & \NVall              & $<6.1$                & ---            \\
  \hline
   \multirow{5}{*}{G395M}
    & \NeIVall            & $<1.2$                & ---                 \\
    & \MgIIall            & $<1.3$                & ---                 \\
    & \NeVL               & $<2.9$                & ---                 \\
    & \OIIall             & $<1.7$                & ---                 \\
    & \NeIIIL             & $<2.7$                & ---                 \\
  \hline
    \end{tabular}
    \caption{Spectral measurements from the three dispersers. Fluxes are in
    units of \fluxcgs, EWs are in (rest-frame) \AA. All uncertainties and
    upper limits are one standard deviation. Empty entries in the EW column are due
    to weak or non-detected continuum. The only secure detection, \CIII,
    is highlighted in bold.
    Fluxes in round parentheses (\OIII, \OII and \NeIII) are considered not
    detected (\S~\ref{s.ne3o2}). All uncertainties use the conservative uncertainties
    from the data reduction pipeline (a discussion of alternative estimates of the
    uncertainties is given in Appendix~\ref{a.data}).\newline
    $^\ddag$ Lines modelled simultaneously across
    the prism and G395M grating. All other lines are modelled as an unresolved Gaussian,
    centred around the redshift of \CIII.
    }\label{t.emlines}
\end{table}

\section{Spectral modelling with \beagle}\label{s.stellar}

We fit the prism spectrum with the Bayesian tool \beagle \citep{Chevallard16}, masking all wavelengths bluer than
1.8332~\mum, to avoid the region around \Lyalpha.
We assume an upper-mass cut-off of the IMF of 300 M$_\odot$ and a delayed exponential SFH, with the last 10~Myr duration set to a constant SFR (which is
free to vary, thus decoupling the present SFR from the previous star formation which mainly contributes to \mstar). We tie the metallicity of stars older than 10~Myr to that of the younger stars and of the nebular gas, and employ the \citet{Charlot+Fall} dust attenuation 
prescription with the fractional attenuation due to the ISM set to 0.4. Specifically, the effective dust $V$-band optical depth \beagletauv is related to the ISM $V$-band optical depth
by $\tau_{V,\mathrm{ISM}} = 0.4 \cdot \beagletauv$.

We find that the observed spectrum is best reproduced with \CO=0.15~\dex (1.4$\times$ solar). We obtain constraints on $\mstar = 7.64\pm0.15$~\MSun, SFR$=1.15\pm0.15$~\MSun~\peryr, and $\beagletauv = 0.12\pm0.04$. The fit indicates a gas-phase oxygen abundance of
$\logoh=7.9\pm0.2$. This value is primarily constrained by
the (presence and absence) of emission lines. As a test, we run a different fit leaving the 
metallicity of stars older than 10~Myr as an 
additional free parameter, but the data are unable 
to constrain this parameter.

From the \beagle stellar
mass, assuming virial equilibrium and the empirical relation of
\citet{cappellari+2013a}, we infer
a second moment of the velocity distribution
$\sigma \approx 20$~\kms, validating our
choice of assuming all emission lines to
be spectrally unresolved. This does not
take into account gas and dark matter,
but even assuming a total mass ten times 
larger than \mstar, we would obtain
$\sigma \approx 65$~\kms, still below the
spectral resolution of our data.

We note that the point-source corrections are optimised for a 5-pixel extraction box.
Yet comparing these results with a 5-pixel extraction with optimal path-loss corrections shows only a marginally higher \mstar and SFR, and \logoh estimates within the 1-\textsigma
uncertainties.
The dust attenuation is, however, inferred to be approximately 2$\times$ higher when fitting 
to a 5-pixel extracted spectrum, suggesting that the slope is somewhat more blue in this
3-pixel extraction (as expected as more flux is lost at longer wavelengths).

\section{A large reservoir of cold and metal poor gas}\label{s.coldgas}

\begin{figure}
    \includegraphics[width=\columnwidth]{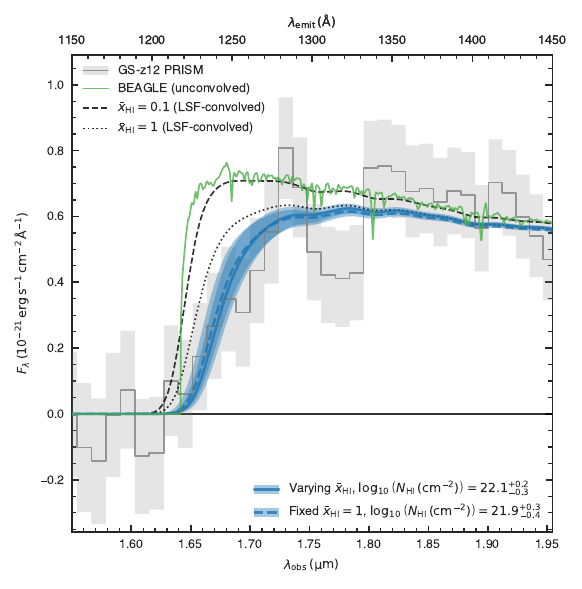}
    \caption{
    Fiducial (maximum-a-posteriori) \beagle spectrum (green) and data (grey) around the \Lyalpha transition. The dashed black line shows the effect of convolving the model spectrum with the prism LSF. The best-fit absorption profiles combining IGM and DLA absorption are shown by the blue lines and blue shading, assuming either a varying IGM neutral hydrogen fraction (solid line) or a fixed fraction $\bar{x}_{\HI} = 1$ (dashed line). The IGM neutral hydrogen fraction is not well constrained, mainly because significant additional absorption is required to reconcile the profiles with the observed spectral break even with maximum IGM absorption ($\bar{x}_{\HI} = 1$; see text for details).}\label{f.dla}
\end{figure}

By fixing the redshift to $z=\zspec$, we can also study the
\Lyalpha absorption profile in more detail. The clear shift between the rest-frame
wavelength of \Lyalpha and the \Lyalpha drop implies a very strong damping-wing absorption, well beyond what is expected from the neutral intergalactic medium (IGM) at such high redshifts \citep[see e.g.][]{heintz+2023a}. The observed profile can be well explained in terms of \Lyalpha absorption in the ISM of the galaxy, or in its surrounding (cold) circum-galactic medium (CGM), similarly to what is observed in  DLA 
systems (Fig.~\ref{f.dla}).  To study the neutral hydrogen absorption in more detail, we consider the maximum-a-posteriori \beagle solution for the stellar continuum, where we have fitted the prism spectrum while masking the region around \Lyalpha (we only used wavelengths above $1.8332 \, \mum$). We then use this fiducial intrinsic spectrum in the Bayesian inference method described below.

First, we attenuate the \beagle spectrum with the absorption profile of a DLA system parametrised only by the column density of neutral hydrogen, \NHI.\footnote{We assume that the DLA system is centred on the systemic redshift of the galaxy, as the spectral resolution at the observed wavelength of \Lyalpha ($\sim 4000 \, \kms$) does not allow us to constrain the infall velocities of gas in the CGM
For the same reason, we note that a pure IGM absorption profile (Fig.~\ref{f.dla}) would require an unphysically large infall velocity to resolve the significant discrepancy with the observed spectrum.} Second, we apply an additional absorption profile arising from neutral gas in the intervening IGM. This is characterised by a (global) neutral hydrogen fraction $\bar{x}_{\HI}$ \citep[under the standard assumption that the gas has mean cosmic density and $T = 1 \, \mathrm{K}$; for more details, see][]{witstok+2023}. The temperature
of the IGM does not impact our results, nor does the temperature of the DLA (we used 100~K
as default, but also tested 1~K and 10,000~K finding no difference in the absorption profile).
We compute the likelihood based on the inverse-variance weighted squared residuals between a given model convolved by the effective LSF (\S~\ref{s.dran}) and the observed prism spectrum.
The logarithm of neutral hydrogen column density, $\log_{10} \NHI \, (\mathrm{cm^2})$, is allowed to vary under a uniform prior distribution between values of $20$ and $23$. We perform the fitting procedure twice, first assuming a uniform prior on $0 \leq \bar{x}_{\HI} \leq 1$ for the IGM neutral hydrogen fraction, and secondly fixing $\bar{x}_{\HI} = 1$. The second scenario allows us to obtain a conservative lower limit on the neutral hydrogen column density in the ISM and/or CGM that is required in addition to IGM absorption to explain the significantly softened spectral break, as seen in Fig.~\ref{f.dla}.

By marginalising over $\bar{x}_{\HI}$, we estimate a neutral hydrogen column density $\log_{10} \NHI \, (\mathrm{cm^2}) = 22.1^{+0.2}_{-0.3}$, while in the
model with fixed $\bar{x}_{\HI} = 1$, we find $\log_{10} \NHI \, (\mathrm{cm^2}) = 21.9^{+0.3}_{-0.4}$.
Both cases clearly require the presence of dense neutral gas in or around \target, as opposed to only the diffuse IGM. In the following, we use the first estimate as our
fiducial value, because it fully considers the degeneracy between the ISM/CGM and the IGM neutral fraction.

Assuming that the dust attenuation through this neutral medium is the same as
through the ISM of \target, i.e. $\tauv=0.05\pm0.02$ (obtained as $0.4\cdot\beagletauv$,
see \S~\ref{s.stellar}), we can obtain an
order-of-magnitude estimate of the metallicity of the DLA gas as
\begin{equation}
    Z_\mathrm{DLA} = Z_\mathrm{MW} \frac{\left( \tauv / N_{\HI}\right)_\mathrm{DLA}}{\left( \tauv / N_{\HI}\right)_\mathrm{MW}} \cdot \frac{\xi_\mathrm{d,MW}}{\xi_\mathrm{d,DLA}}.
\end{equation}

The gas-to-extinction ratio we observe in the Milky Way is
$\NHI/\tauv=(1.92\pm0.02)\times 10^{21}\,\percm{-2}$ \citep[e.g.,][]{zhu+2017}, and the dust-to-metal ratio is $\xi_\mathrm{d,MW}=0.45$
\citep[e.g.,][]{konstantopoulou+2023}.
The metallicity of the ISM in the solar neighbourhood is 0.2~dex lower than 
solar \citep{arellano-cordova+2021}, therefore, we assume $Z_\mathrm{MW} = 0.6$ solar.
At low metallicity and at high gas fractions (as appropriate for \target) \xid is lower \citep{Decia16,devis+2019}. Assuming a metallicity in the range
0.03--0.1 solar, the typical \xid measured in high-redshift
absorbers is $0.24\pm0.11$\footnote{We took the median of DLAs illuminated by GRBs, but taking the median of
DLAs illuminated by QSOs would change the \xid by a factor of $\approx 2$,
which is still within the large scatter of these measurements.} \citep[e.g.,][]{konstantopoulou+2023}.
These values are on average half the
Milky Way value. The above formula then gives us a DLA metallicity
of 0.004--0.02 solar ($\logoh=6.3\text{--}7.0~\dex$), much lower than the estimate from \S~\ref{s.stellar}.
We note that the large uncertainties in our assumptions are compounded by systematics in
the spectral modelling with \beagle. For instance,
using a 5-pixel extraction window (instead of the default 3-pixel) increases
\tauv by approximately a factor 2.
In addition, the optical depth \tauv estimated with \beagle relies on an `attenuation' law (including absorption and scattering into and out of the line of sight caused by local and global geometric effects), while the column density from the DLA-like fit assumes a pure foreground `extinction' curve (including absorption and scattering only out of the line of sight). Adopting an extinction rather than an attenuation curve in \beagle would provide lower \tauv -- which would lower the metallicity estimate.
Therefore, in the following, we treat
our mean metallicity estimate of 6.6~dex as an upper limit.

Assuming all this gas is from the galaxy's ISM -- and that the DLA comes
mostly from within one \re -- we can estimate the mass of atomic gas as $\mgas =
2~(1.34~m_\mathrm{H}~\NHI)~\text{\textpi} q \re^2$,
giving
$\log\,(\mgas/\MSun) = 7.0\pm0.25$
(where $m_\mathrm{H}$ is the
mass of the hydrogen atom, the factor 1.34 accounts for the helium fraction, the additional
factor of 2 is a geometric factor assuming equal
column density on the far side of the galaxy,
and $q$ is the
axis ratio of the galaxy; we used the morphological parameters from Table~\ref{t.physpar}). This gas mass is 10--50~per cent of the stellar mass inferred by \beagle. 
However, this is only the mass of atomic hydrogen and does not take into account the fraction of molecular (or ionised) hydrogen, hence it should be considered a lower limit.
Such a gas fraction (or even higher given that it is a lower boundary) is consistent with many other galaxies at high redshift \citep{tacconi+2020}.

The corresponding lower limit on the gas surface density is 150~\MSun~pc$^{-2}$ -- at the boundary between normal
star-forming regions and starbursts galaxies in the local
Universe \citep[e.g.,][]{kennicutt+evans2012}. At these densities, most
cold gas in the local Universe is expected to be in the molecular phase.

The SFR density inferred combining \beagle and the size measurement is
of order $100~\MSun~\peryr~\mathrm{kpc}^{-2}$. Together with the gas surface density estimated above, this would place this galaxy well above the Schmidt--Kennicutt (S-K) relation, even for starburst galaxies. However, it could be consistent with the S-K relation considering that the surface density inferred above is a lower limit, and implying, once again, that a significant fraction of the gas is in molecular form.

\section{Source of photo-ionisation and chemical abundances}\label{s.diagn}

\begin{figure}
   \includegraphics[width=\columnwidth]{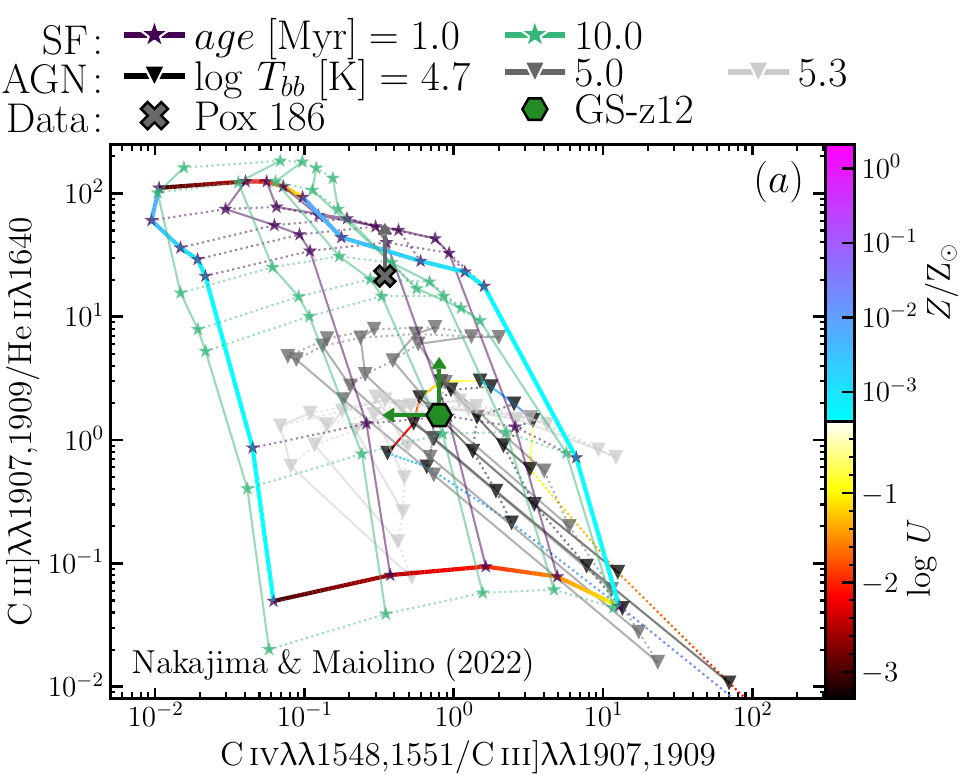}
   \includegraphics[width=\columnwidth]{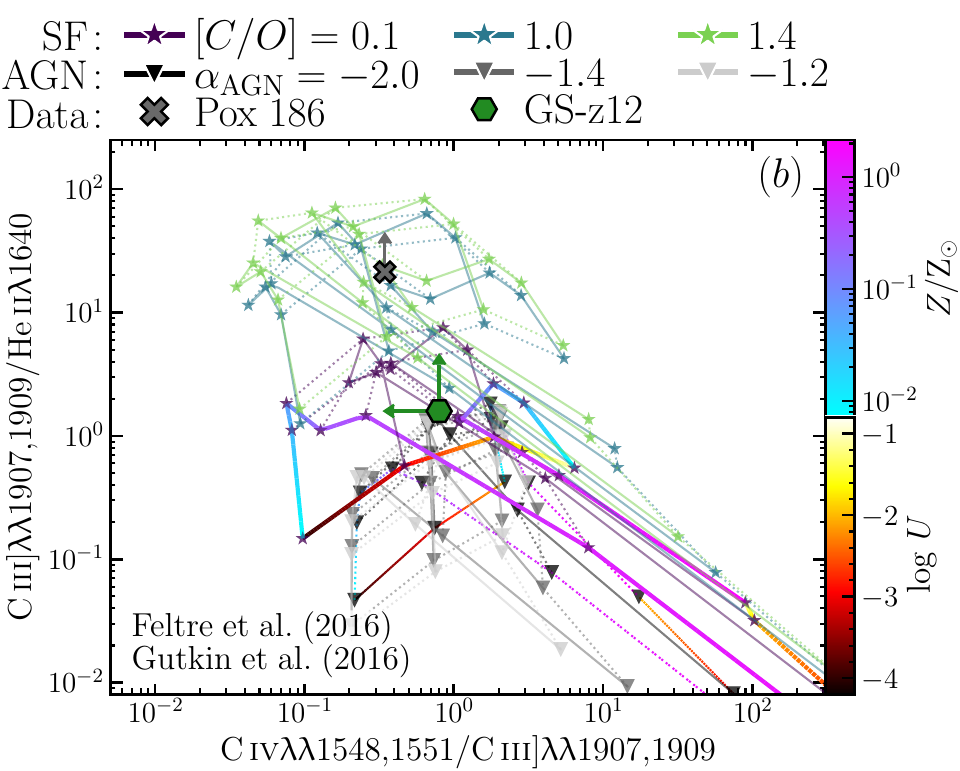}
   \includegraphics[width=\columnwidth]{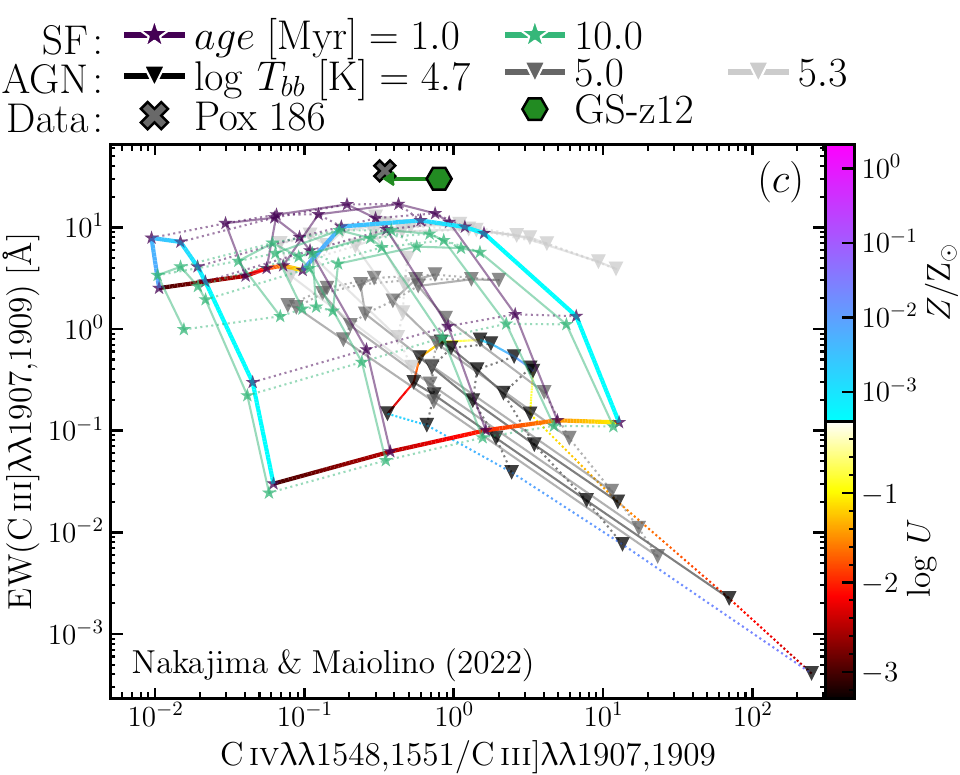}
   {\phantomsubcaption\label{f.diag.a}
    \phantomsubcaption\label{f.diag.b}
    \phantomsubcaption\label{f.diag.c}
   }
   \vspace{-1em}
   \caption{Panels~\subref{f.diag.a}--\subref{f.diag.b}; photo-ionisation diagnostics for star-formation (stars) and AGN (triangles),
   compared to \target (hexagon) and to the local analogue \pox
   \citetext{cross, \citealp{kumari+2023}}.
   Panel~\subref{f.diag.c}; photo-ionisation diagnostic with \EW{\CIII}.
   The grids are from \citetalias{nakajima+maiolino2022}
   (panels~\subref{f.diag.a} and \subref{f.diag.c}), and from
   \citetalias{gutkin+2016} and \citetalias{feltre+2016}
   (panel~\subref{f.diag.b}). All upper and lower limits are given at 3~\textsigma.
   }
   \label{f.diag}
\end{figure}

\begin{figure}[!ht]
   \includegraphics[width=\columnwidth]{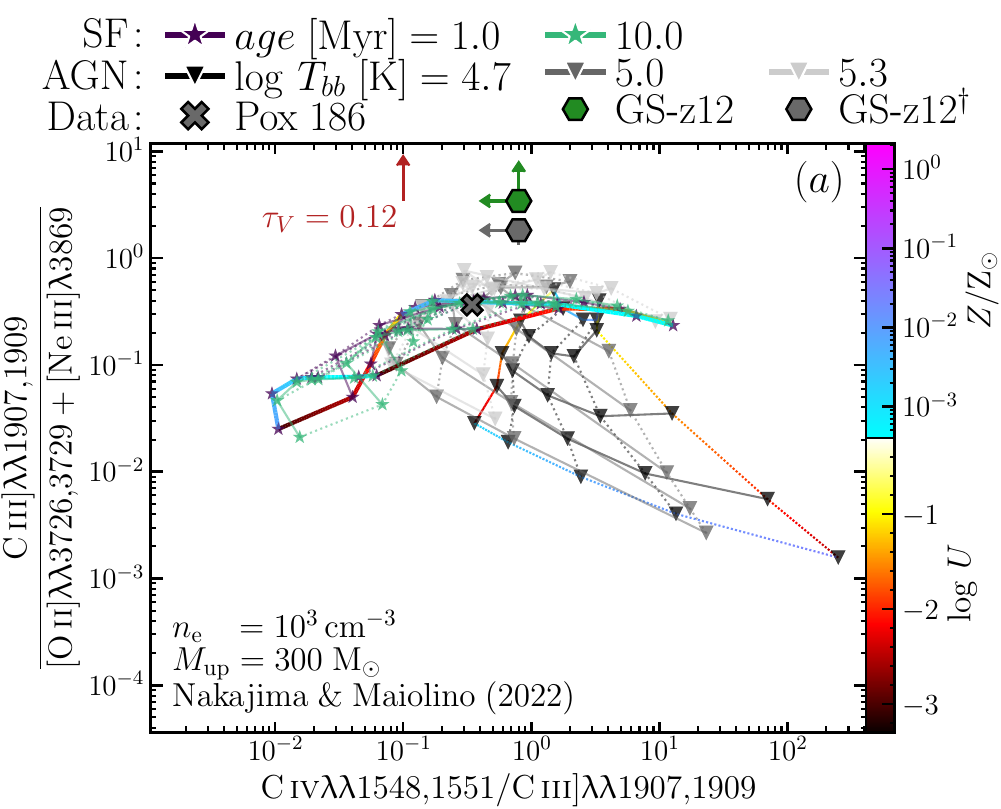}
   \includegraphics[width=\columnwidth]{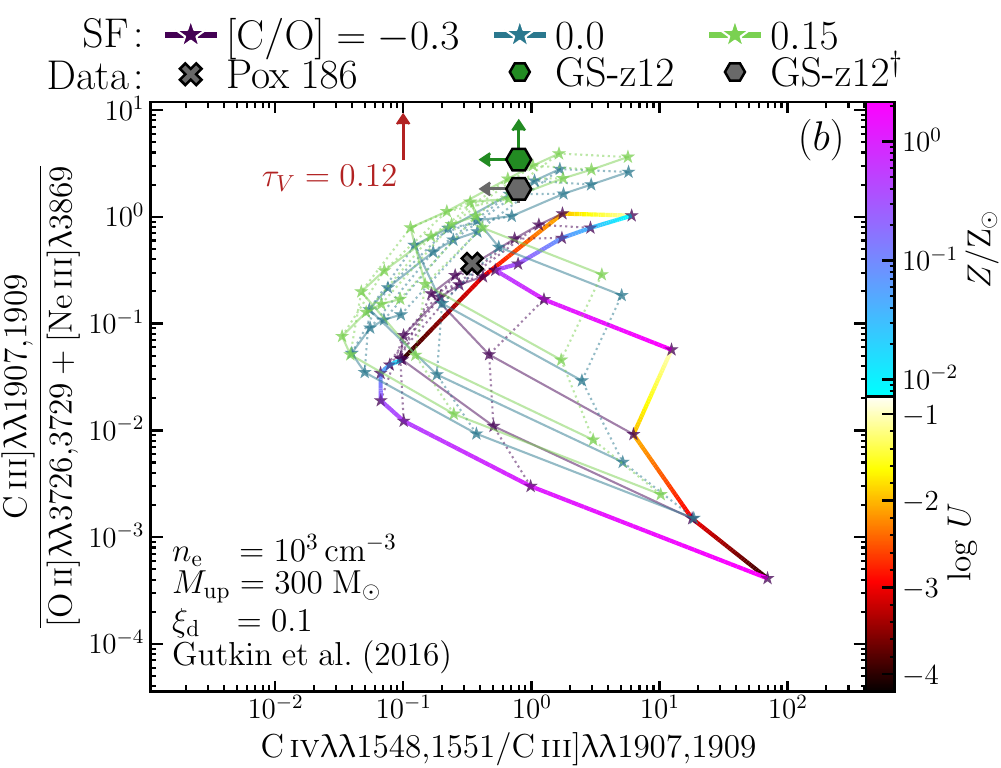}
   {\phantomsubcaption\label{f.coab.a}
    \phantomsubcaption\label{f.coab.b}
   }
   \caption{Panel~\subref{f.coab.a}; our upper limits on \target
   (hexagon), compared to photo-ionisation models
   powered by star formation (SF; stars) and AGN
   \citepalias[triangles;][]{nakajima+maiolino2022}, illustrating
   that \target lies well outside the range of grids with
   solar and subsolar C/O \citetext{used by \citetalias{nakajima+maiolino2022}}.
   Panel~\subref{f.coab.b}; star-forming models
   of \citetalias{gutkin+2016}, with varying C/O abundance between
   0.5 and 1.4 solar. Combining the detection of \CIIIL with
   stringent upper limits on \OIIall and \NeIIIL, the
   models constrain the abundance
   to super-solar values, i.e. much higher values than what is observed in
   galaxies at $z=9$ (see Fig.~\ref{f.cooh}) and in the local analogue
   \pox \citep{kumari+2023}. Alternatively, assuming both \OII and \NeIII to be detected, we obtain the grey hexagon marked with a $^\dagger$, which points to a solar \CO\ -- still high for high-redshift galaxies.
   The red
   upward arrow marks the magnitude of the dust reddening
   correction, which we conservatively \textit{do not 
   apply.} We colour code only the external
   lines in the grids, and only for one of the SF and AGN grids;
   note the different range of the
   grid parameters.
   }
   \label{f.coab}
\end{figure}

\begin{figure}
   \includegraphics[width=\columnwidth]{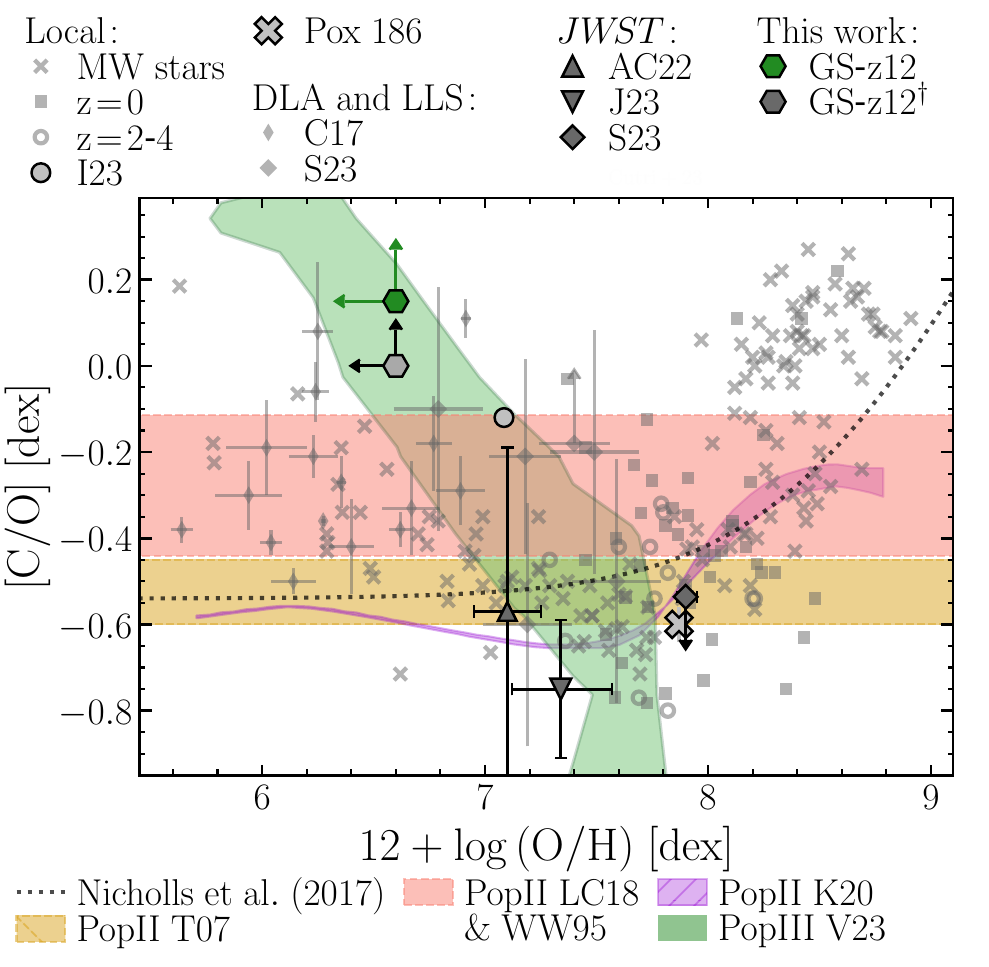}
   \caption{Compilation of chemical abundance patterns from both local stars and galaxies
   (see text for references) and from metal-poor galaxies at $z=2\text{--}9$,
   compared to the yields of PopII and PopIII
   supernovae. \target stands out as a metal-poor system with high C/O abundance, and is therefore
   closer to DLA systems
   \citetext{\citealp{cooke+2017}, C17; \citealp{saccardi+2023}, S23}
   than to other high-redshift galaxies recently observed by \jwst
   \citetext{\citealp{arellano-cordova+2022}, AC22; \citealp{jones+2023}, J23;
   \citealp{stiavelli+2023}, S23}.
   The dotted line is the
   chemical enrichment pattern from \citet{nicholls+2017};
   the golden horizontal hatched band is the yield
   of PopII supernovae from \citet{tominaga+2007}. 
   The purple region is
   the PopII abundance pattern from
   \citet{kobayashi+2020}.
   The red
   horizontal band is the combined yield of 
   metal-poor Type-II supernovae from \citet{woosley+weaver1995} and \citet{limongi+chieffi2018}. The green region is
   the yield of PopIII supernovae
   \citetext{\citealp{heger+woosley2010}, adapted from \citealp{vanni+2023}}.
   We adopt the solar abundance ratio $\log (\mathrm{C/O})
   = -0.26$ \citep{asplund+2009}.
   }
   \label{f.cooh}
\end{figure}
To identify the source of photo-ionisation in \target, we rely on models
from the literature. The high value of \EW{\CIII} rules out pristine
sources like Population~III stars and direct-collapse black holes
\citetext{\citealp{nakajima+maiolino2022}; hereafter: \citetalias{nakajima+maiolino2022}}.
We therefore focus on star-formation
and AGN photo-ionisation, using models from \citetalias{nakajima+maiolino2022}, \citet[][hereafter: \citetalias{feltre+2016}]{feltre+2016} and
\citet[][hereafter: \citetalias{gutkin+2016}]{gutkin+2016}. In the following, the grids of star-forming models 
are marked with stars, and those of AGNs are marked by triangles.
For the star-forming models of \citetalias{nakajima+maiolino2022}, we
fix the upper-mass cut-off of the IMF to 300~\MSun; for their AGN models, we fix the 
shape of the ionising continuum to have a slope $\alpha_\mathrm{AGN}=-2$.
In both cases, the density is $\nelec=10^3\,\percm{-3}$.
For the star-forming models of \citetalias{gutkin+2016}, we fix the upper-mass cut-off of the IMF to 300~\MSun, the dust-to-metal mass ratio $\xid=0.1$, the
maximum stellar age to 100~Myr, and the density to $\nelec=10^2\,\percm{-3}$. 
For the AGN models of
\citetalias{feltre+2016}, we fix both $\xid=0.1$ and
$\nelec=10^3\,\percm{-3}$.
We colour-code the external envelope of some of the grids by their
metallicity $Z$ and ionisation parameter $\log\,U$.
Each of these models presents unique advantages:
\citetalias{nakajima+maiolino2022} extend to very low metallicity, and
also provide EWs; the other two models provide independently varying C/O abundances.

In Fig.~\ref{f.diag.a}--\subref{f.diag.b} the green hexagon shows the NIRSpec 
3-\textsigma upper limits in the UV diagnostic diagram using \CIV, 
\CIII and \HeIIL
\citepalias[hereafter: \HeII;][]{feltre+2016,gutkin+2016}.
We also report the position of the local analogue \pox as a cross; while the stringent upper limits on
\pox place it confidently in the star-forming region of the diagram,
\target is in the upper envelope of the AGN models
(panel~\subref{f.diag.a}). Given the trend of increasing
\CIII/\HeII with decreasing metallicity, we expect that extending
the metallicity range of the \citetalias{feltre+2016} grid to lower values would also allow these
models to overlap with \target.
In Fig.~\ref{f.diag.c}, we show \EW{\CIII} vs 
\CIV/\CIII for the \citetalias{nakajima+maiolino2022} models;
both \target and the local analogue \pox cannot be explained
by either star-formation or AGN photo-ionisation. However, we note
that the \citetalias{nakajima+maiolino2022} models assume the C/O
enrichment scaling derived from the local Universe \citep{dopita+2006};
therefore, at the lowest metallicities relevant to this work, their
C/O abundance is significantly sub-solar, and actually closer to the
pure yield of Type-II supernovae.\\

In Fig.~\ref{f.coab} we explore the C/O abundance ratio using the
upper limits on both \OII and \NeIII -- leveraging the fact that
both O and Ne are \textalpha elements released by Type-II supernovae.
The standard enrichment pattern at the root of the models of
\citetalias{nakajima+maiolino2022} cannot
explain the observed lower limit (panel~\subref{f.coab.a}).
The star-forming models of \citetalias{gutkin+2016}, in contrast, allow C/O to
vary independently of the total metallicity.
To explain the observations of \target (green hexagon), we thus require
a super-solar $\CO>0.15~\dex$ \citetext{i.e., 
higher than the highest value explored by the \citetalias{gutkin+2016} models}. Even assuming
\OII and \NeIII detections, we would still 
obtain $\CO\approx0~\dex$ (grey hexagon).
This abundance
pattern, with higher-than-solar C/O, would also alleviate the
tension with the EW measurements (Fig.~\ref{f.diag.c}).
Changing the upper-mass cut of the IMF from 300 to 100~\MSun and the maximum stellar age from 100 to 10~Myr
does not change our conclusions.
Increasing the dust-to-metal mass fraction \xid from 0.1 to 0.3 would instead make our conclusions even stronger, by
lowering the predicted
\CIII/(\OII+\NeIII) ratio
(because C has a higher depletion than Ne and O).

Fig.~\ref{f.cooh} shows the abundance ratio C/O vs \logoh, including
Milky Way stars
\citep[grey crosses;][]{gustafsson+1999,akerman+2004,fabbian+2009,nissen+2014},
galaxies ranging from the local Universe
\citep[squares, labelled $z=0$;][]{berg+2016,pena-guerrero+2017,senchyna+2017,berg+2019}
up to Cosmic Noon \citep[empty circles, labelled $z=2$-4;][]{
erb+2010,christensen+2012,bayliss+2014,james+2014,stark+2014,berg+2018,mainali+2020,matthee+2021,iani+2023},
and DLAs \citep{cooke+2017,saccardi+2023}. We also show the abundances (and upper limits)
from four high-redshift galaxies at $z=6\text{--}9$ measured by \jwst
\citep{arellano-cordova+2022,jones+2023,stiavelli+2023}. These galaxies tend to display a low
[C/O]$\approx-0.6\,\dex$, very close to 
or even lower than the yield of metal-poor
Type-II supernovae \citetext{golden horizontal
band, \citealp{tominaga+2007}; red horizontal band, \citealp{woosley+weaver1995} and \citealp{limongi+chieffi2018}; purple
region, \citealp{kobayashi+2020}}.
Indeed, these
galaxies follow qualitatively the enrichment sequence from \citet{nicholls+2017},
which is interpreted as a mixing sequence between the pure Type-II yields of
young, low-metallicity systems, and the later contribution of stars on the
asymptotic giant branch (AGB), which have longer enrichment timescales than Type-II
supernovae \citep[e.g.,][]{salaris+2014}.

\target, instead, appears as an outlier relative to other galaxies, but it is consistent with some high-z DLAs. Specifically, its \logoh metallicity has an upper limit of 6.6~\dex, as estimated 
from combining \NHI and \tauv (\S\S~\ref{s.stellar} and~\ref{s.coldgas}).
The C/O abundance is a lower limit, based on the lower limit on the
\CIII/(\OII+\NeIII) line ratio (\S~\ref{s.diagn}). These values place \target
clearly outside the rising enrichment sequence of local and lower redshift galaxies, and closer to the
decreasing branch occupied by DLAs and extremely metal-poor stars in the Milky Way halo.

When compared with the models discussed above, the chemical enrichment pattern of \target is inconsistent with pure Type-II supernovae yields.
Yet, its C/O and low metallicity suggest an enrichment history more similar to
DLAs (small grey triangles). The yields of PopIII supernovae
from \citet[][green band]{vanni+2023} give a sequence of super-solar C/O that may
explain the lower limit of \target. Crucially, these models produce super-solar C/O
via low-energy PopIII supernovae 
\citep[$E<2\times 10^{51}$~erg; ][]{vanni+2023}.

\section{Summary, discussion and outlook}\label{s.discon}

\jwst NIRSpec MSA observations from the combined programmes PID~1210 and 3215 enabled 
us to investigate the
detailed physical properties of \target, a galaxy at $z>12$, near Cosmic Dawn.
We report the detection of \CIII at $z=\zspec$. This is the most distant nebular-line detection to date, and the most distant evidence of chemical enrichment.

A full spectral modelling with the \beagle tool confirms the earlier results from
\citetalias{curtis-lake+2023}; \target has a stellar mass $\mstar = 4\times10^7~\MSun$
and a mass-doubling time of 40~Myr. The fit indicates a moderate dust attenuation optical depth,
$\tauv = 0.12$, and a sub-solar metallicity of \logoh=7.9~\dex, mostly constrained by the emission lines.

The redshift obtained through the \CIII line implies that the \Lyalpha
drop has a  prominent damping wing. This
cannot be associated only with IGM
absorption, but can be modelled with absorption by
the ISM of the galaxy or its CGM (i.e., a local
DLA). Specifically, thanks to the accurate redshift obtained from the nebular emission lines, we can reliably model the damping wing profile and infer a high column density of
$\NHI \approx 10^{22}~\percm{-2}$.
The inferred gas fraction (\mgas/\mstar) is about 0.1--0.5, consistent with what one
would expect at these high redshifts \citep{carilli+walter2013,tacconi+2020}.
We note, however, that the uncertainties on
 the conversion from \NHI to \mgas and on the stellar mass-to-light ratio 
\citep[due for instance to a top-heavy IMF, e.g.,][]{rusakov+2023} can be large. Converting this gas mass and the \beagle SFR into surface densities,
we obtain values 150~\MSun~pc$^{-2}$ and 100~\MSun~\peryr~kpc$^{-2}$,
respectively. This combination is one to two orders of magnitude higher than the predictions from the local Schmidt--Kennicutt relation \citep{kennicutt+evans2012}, suggesting that we are missing a substantial
fraction of the gas, very likely in the molecular form, or that our SFRs are significantly overestimated.

The gas metallicity we infer from \NHI is also quite uncertain, given 
the unwarranted assumption that \tauv estimated from the
stellar continuum also applies to the \Lyalpha absorber. This is compounded by
the large uncertainties on the dust-to-metal ratio (\xid) at
these early epochs. With an average estimate of $\xid=0.24$
\citep{konstantopoulou+2023} we obtain a sub-solar metallicity of about 0.004--0.02 solar, smaller than
the value inferred from \beagle.

The high \EW{\CIII} is reminiscent of lower-redshift AGN \citep{lefevre+2019},
but the lack of \CIV emission seems at odds with this scenario. Unfortunately, and in spite of a 50 h-long integration, the upper limits on \CIV and \HeII are 
unable to definitely rule out the presence of an AGN (Fig.~\ref{f.diag.a}--\subref{f.diag.b}; 
we set a 
threshold of 3-\textsigma for the non detections, or 99.9~per cent confidence).
A local galaxy with properties analogue to \target, i.e. with a high-EW \CIII emission, is the low-mass, metal-poor 
star-forming dwarf galaxy \pox \citep[]{kumari+2023}. \target and \pox have 
similar size of $\approx$100~pc,
yet, they have different masses (by two orders of magnitude) and markedly different C/O abundances.

The detection of \CIII -- and its high EW -- rules out scenarios of pristine stellar populations (PopIII) or black holes \citetext{direct-collapse black holes; \citetalias{nakajima+maiolino2022}}.
The high-EW value of $30\pm7~\AA$ cannot be explained with
the sub-solar C/O ratio of the \citetalias{nakajima+maiolino2022} models. 
\citet{kumari+2023} have
found similar difficulties in explaining the emission-line properties of 
\pox (grey cross in Fig.~\ref{f.diag.c}) -- perhaps exacerbated by the
sub-solar C/O measured in their dwarf galaxy.

The long integration time of these observations provides stringent upper limits
on both \OII and \NeIII, which we use to study the C/O ratio 
(Fig.~\ref{f.coab}).
The inability of subsolar- C/O models to explain the
observed lower limits on \CIII/(\OII+\NeIII) suggests that the bulk of
the gas in \target has super-solar C/O abundance.
While the prism data indicate a tentative
detection of \OII and \NeIII, we do not consider these detections to be robust (see \S~\ref{s.ne3o2}).
Nevertheless, if we replace the 3-\textsigma upper limits in
\CIII/(\OII+\NeIII) with detections, the
line ratio decreases from 3.4 down to 1.8, which would still 
require a solar C/O (Fig.~\ref{f.coab}).
In addition, throughout our analysis we do not 
apply any dust correction, but we note that any
correction would increase the \CIII/(\OII+\NeIII)
ratio and the C/O abundance, making our result even
stronger.

We note that the lower limit on C/O that we obtain is based on photo-ionisation models, which themselves
rely on several assumptions about the shape of the ionising sources
and the gas chemical abundance patterns. However, taking our results at face value,
we find valuable implications on the chemical evolution history of this early galaxy.

In Fig.~\ref{f.cooh}, we showed the location of \target in the O/H vs C/O
enrichment diagram. The high C/O and low metallicity of this galaxy are inconsistent with the yields of Type-II supernovae
([C/O]$\lesssim -0.15$~dex), and much higher than what found by \jwst in
galaxies at $z=6\text{--}9$ \citep{arellano-cordova+2022,jones+2023,stiavelli+2023}.

Stars with extremely high C abundance have been found in the Milky Way, and
these have some of the lowest measured Fe/H abundances \citep{aoki+2006}.
While for these old galactic stars other enrichment channels are possible
(dredge up or interactions with asymptotic-giant-branch companions, AGB), these 
scenarios seem inadequate for the case of \target, as dredge up is most appropriate
for evolved low-mass stars \citep[as is the case for ][]{aoki+2006}.
Specifically, AGB stars would need to dominate the chemical enrichment history of \target, but at $z=\zspec$ only stars more massive than $3~\MSun$ 
have had sufficient time to reach this phase, implying that the chemical enrichment history is still dominated by core-collapse supernovae. Besides, most metal-poor, C-rich stars in the
Milky Way have solar or even sub-solar C/O abundance \citetext{with \citealp{aoki+2006}
representing the exception, with $\CO>0.26$}.

A possible explanation for this anomalously high \CO value is that the stars in 
\target carry the signature of chemical enrichment due to massive, metal-poor 
stars, whose yields can be
significantly different than those of
PopII Type-II supernovae. In particular, chemical enrichment from PopIII can potentially explain the observed C/O in \target
\citep{heger+woosley2010,limongi+chieffi2018},
especially when including low-energy supernova
explosions \citep{vanni+2023}.
These explosions leave remnants with a higher fraction of the O-rich shell locked in them, with respect to the C-rich shell, thereby
increasing the C/O yield of the ejecta to solar or even super-solar values.
An alternative explanation is represented by peculiar star-formation histories, such as those invoked to explain the abundance patterns in GN-z11
\citep{kobayashi+ferrara2023}. Such peculiar SFHs would, however, also produce a strong nitrogen abundance, which is inconsistent with the observations of \target.

This detection of the most distant metal transition, which has provided such precious information about the earliest phases of the chemical enrichment, has required a very long exposure (65~h, although mostly as a parallel observation). This is due to the extreme faintness of such distant galaxies.
However, in the future, large-area surveys and gravitational lenses may help identify more high-redshift galaxies that are sufficiently bright for deep spectroscopic follow up with shorter exposures.

\begin{acknowledgements}
We are grateful to S. Salvadori and I. Vanni for providing the chemical enrichment tracks for some Population III scenarios. We thank N.~Laporte for useful
discussions and suggestions.
FDE, RM, JW, WB, TJL acknowledge support by the Science and Technology Facilities Council (STFC), by the ERC through Advanced Grant 695671 ``QUENCH'', and by the
UKRI Frontier Research grant RISEandFALL. RM also acknowledges funding from a research professorship from the Royal Society.
SC and GV acknowledge support by European Union's HE ERC Starting Grant No. 101040227 - WINGS.
ECL acknowledges support of an STFC Webb Fellowship (ST/W001438/1).
AJB and JC acknowledge funding from the ``FirstGalaxies'' Advanced Grant from the European Research Council (ERC) under the European Union's Horizon 2020 research and innovation programme (Grant agreement No. 789056).
SA acknowledges support from Grant PID2021-127718NB-I00 funded by the Spanish Ministry of Science and Innovation/State Agency of Research (MICIN/AEI/ 10.13039/501100011033). 
This research is supported in part by the Australian Research Council Centre of Excellence for All Sky Astrophysics in 3 Dimensions (ASTRO 3D), through project number CE170100013.
KH, ZJ, BDJ, MR, BR and CNAW acknowledge support from the JWST/NIRCam Science Team contract to the University of Arizona, NAS5-02015; DJE is also supported as a Simons Investigator.
KN acknowledges support from JSPS KAKENHI Grant JP20K22373.
RS acknowledges support from a STFC Ernest Rutherford Fellowship (ST/S004831/1).
H{\"U} gratefully acknowledges support by the Isaac Newton Trust and by the Kavli Foundation through a Newton-Kavli Junior Fellowship.

This work was performed using resources provided by the Cambridge Service for Data Driven Discovery (CSD3) operated by the University of Cambridge Research Computing Service (\href{www.csd3.cam.ac.uk}{www.csd3.cam.ac.uk}), provided by Dell EMC and Intel using
Tier-2 funding from the Engineering and Physical Sciences Research Council (capital grant EP/T022159/1), and DiRAC funding from the Science and Technology Facilities Council (\href{www.dirac.ac.uk}{www.dirac.ac.uk}).

This work also made extensive use of the freely available
\href{http://www.debian.org}{Debian GNU/Linux} operative system. We used the
\href{http://www.python.org}{Python} programming language
\citep{vanrossum1995}, maintained and distributed by the Python Software
Foundation. We further acknowledge direct use of
{\sc \href{https://pypi.org/project/astropy/}{astropy}} \citep{astropyco+2013},
\beagle \citep{Chevallard16},
{\sc \href{https://pypi.org/project/matplotlib/}{matplotlib}} \citep{hunter2007},
{\sc \href{https://pypi.org/project/numpy/}{numpy}} \citep{harris+2020},
{\sc \href{https://scikit-learn.org/stable/index.html}{scikit-learn}} \citep{scikit-learn},
{\sc \href{https://pypi.org/project/scipy/}{scipy}} \citep{jones+2001}
and {\sc \href{http://www.star.bris.ac.uk/~mbt/topcat/}{topcat}} \citep{taylor2005}.
\end{acknowledgements}

\appendix
\section{In-depth analysis of the data}\label{a.data}

\begin{figure*}
  \includegraphics[width=\textwidth]{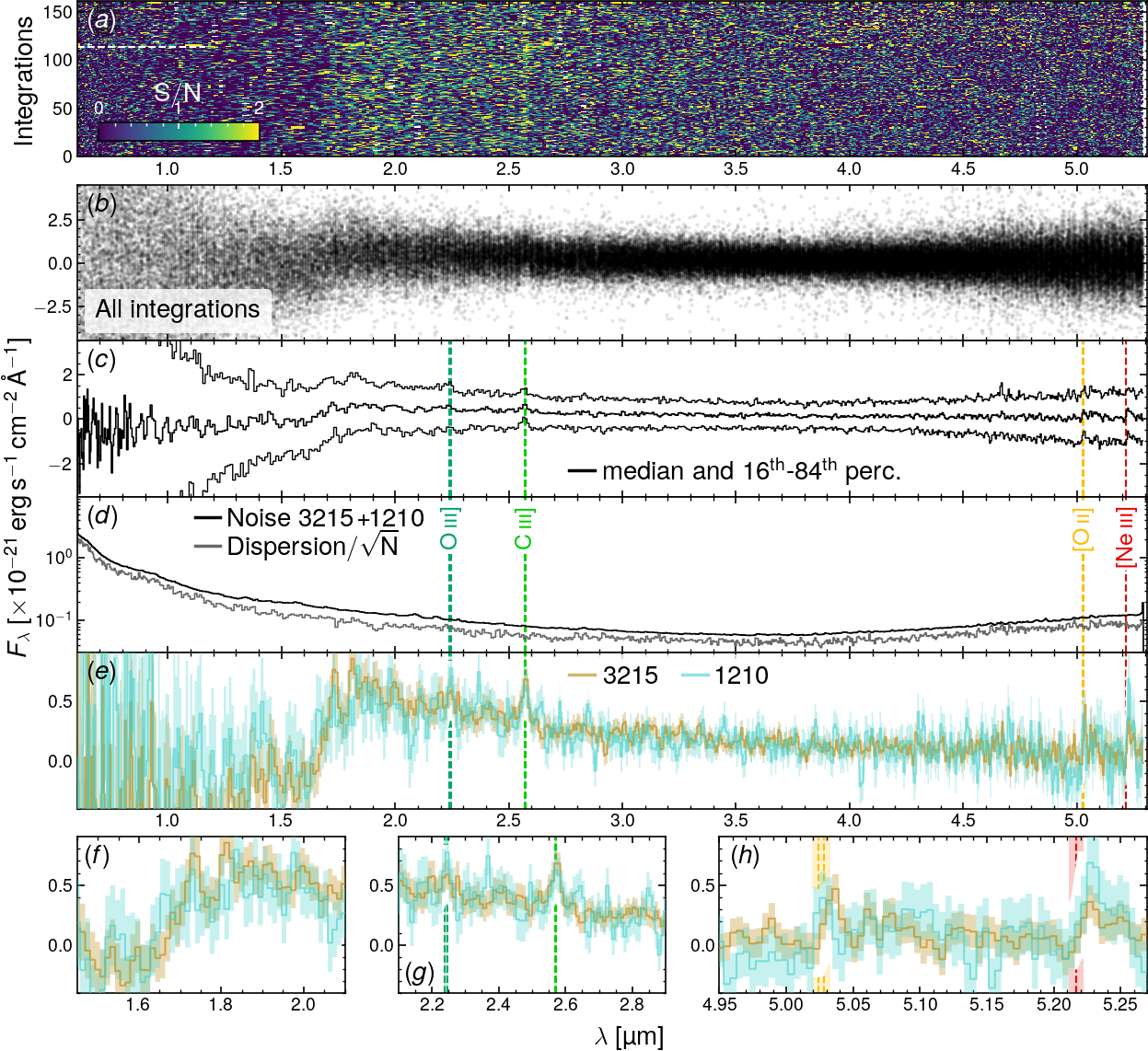}
  {\phantomsubcaption\label{f.integs.a}
   \phantomsubcaption\label{f.integs.b}
   \phantomsubcaption\label{f.integs.c}
   \phantomsubcaption\label{f.integs.d}
   \phantomsubcaption\label{f.integs.e}
   \phantomsubcaption\label{f.integs.f}
   \phantomsubcaption\label{f.integs.g}
   \phantomsubcaption\label{f.integs.h}
  }
  \caption{Summary of the prism observations.
  Panel~\subref{f.integs.a}; S/N of each
  of the 162 individual integrations. The horizontal dashed line separates the bottom
  114 rows  with the 3215 integrations from the
  1210 integrations above.
  Panel~\subref{f.integs.b}; spectrum of each of
  the individual integrations. A feature
  at the location of \CIII is clearly seen.
  Panel~\subref{f.integs.c}; median and
  16\textsuperscript{th}--84\textsuperscript{th}
  percentile of panel~\subref{f.integs.b}, showing
  again the presence of \CIII, and the possible
  presence of \OII.
  Panel~\subref{f.integs.d}; comparison between the
  noise estimated by the data reduction pipeline (black)
  with the noise estimated from the dispersion of the
  integrations (grey, calculated under a Gaussian-noise
  assumption).
  Panels~\subref{f.integs.e}--\subref{f.integs.h};
  comparison between 3215 (sand) and 1210 (cyan),
  showing that \CIII is seen in both
  programmes; in contrast, \OIII is not seen in
  3215 (the deepest of the two surveys); an emission line close to the
  position of \OII is
  seen in both surveys, but at different
  wavelengths; \NeIII is seen in 3215, but has
  an unconvincing line profile in 1210.
  }\label{f.integs}
\end{figure*}

In \S\S~\ref{s.c3} and~\ref{s.ne3o2} we summarised our considerations on the
reliability of the emission lines detections. The goal of this section is to
provide more information on the data, and on the reliability of the various
emission-line detections -- particularly in the low-S/N regime.
To do so, we take advantage of 162 individual integrations for \target, 
each with 1,400.5~s duration. We show the S/N as a function of wavelength
for each of these detections in Fig.~\ref{f.integs.a}; the y-axis values
from 0 to 113 correspond to the 114 integrations five visits from
PID~3215; the remaining 48 integrations are from two visits in PID~1210. We saturate the colour limits between
S/N 0 and 2, to highlight a faint vertical line at 2.57~\mum, which we
interpret as \CIII. No other emission lines are clearly visible in
this representation. In particular, we see no evidence for \OII or
\NeIII; even though these two lines have lover S/N than \CIII (which
should make them harder too see), they occupy a region of the spectrum with lower continuum flux than \CIII (which would enhance
the contrast with neighbouring pixels).
The 162 individual spectra are also shown in
panel~\subref{f.integs.b}; while this figure is dominated by noise, the
\CIII line is also visible here.
Panel~\subref{f.integs.c} shows the 16\textsuperscript{th}, 50\textsuperscript{th} and 84\textsuperscript{th} percentiles of the data
in panel~\subref{f.integs.b}. The vertical lines mark the locations of
\OIII, \CIII, \OII and \NeIII. In this figure, \CIII and \OII are clearly
seen, while \OIII seems absent and so does \NeIII. This line in
particular has an ugly spectral profile, which further validates our decision to treat it as a tentative detection only.

Panel~\subref{f.integs.d} shows a comparison between the noise of our
co-added data, derived from the data reduction pipeline (black) and the
dispersion of the 162 exposures (grey), estimated as the 
84\textsuperscript{th}-16\textsuperscript{th} inter-percentile 
range divided by $2\cdot\sqrt{162}$. The latter estimate is 25--30~per cent
smaller than the former, but it does not take into account correlated noise
due to spectral resampling on the NIRSpec detector. In contrast, the
data reduction pipeline implements an effective correction to the noise
\citep{dorner+2016}, giving a more conservative estimate.
Using the grey curve as noise vector, the \CIII S/N would increase to 7--7.5.

Finally, in panel~\subref{f.integs.e} we show the 1-d
spectra from PID~1210 (cyan) and 3215 (sand), including zoom-in windows
around the \Lyalpha drop, \OIII--\CIII and \OII-\NeIII 
(panels~\subref{f.integs.f}--\subref{f.integs.h}). This comparison in 
particular shows that \OIII seems absent in 3215, but a single high pixel in
1210 may be driving its 2.3 tentative detection; \CIII, in contrast, appears 
to have similar profiles at the same location in both 1210 and 3215 
(panel~\subref{f.integs.g}).
\OII appears both in 1210 and 3215, but, as we have mentioned, its wavelength
is different between the two observations. An issue due to the wavelength
solution seems unlikely, because the galaxy is well centred in the MSA
shutter, and because the shift in spectral pixels due to intra-shutter
position is negligible at these red wavelengths. If any problem in the
wavelength calibration was present, it could potentially explain the moderate
mismatch in redshift between \CIII and \OII--\NeIII.
\NeIII confirms its dubious nature, especially in 1210, where it shows a
narrow profile that does not resemble a true emission line.

\begin{figure*}
  \includegraphics[width=\textwidth]{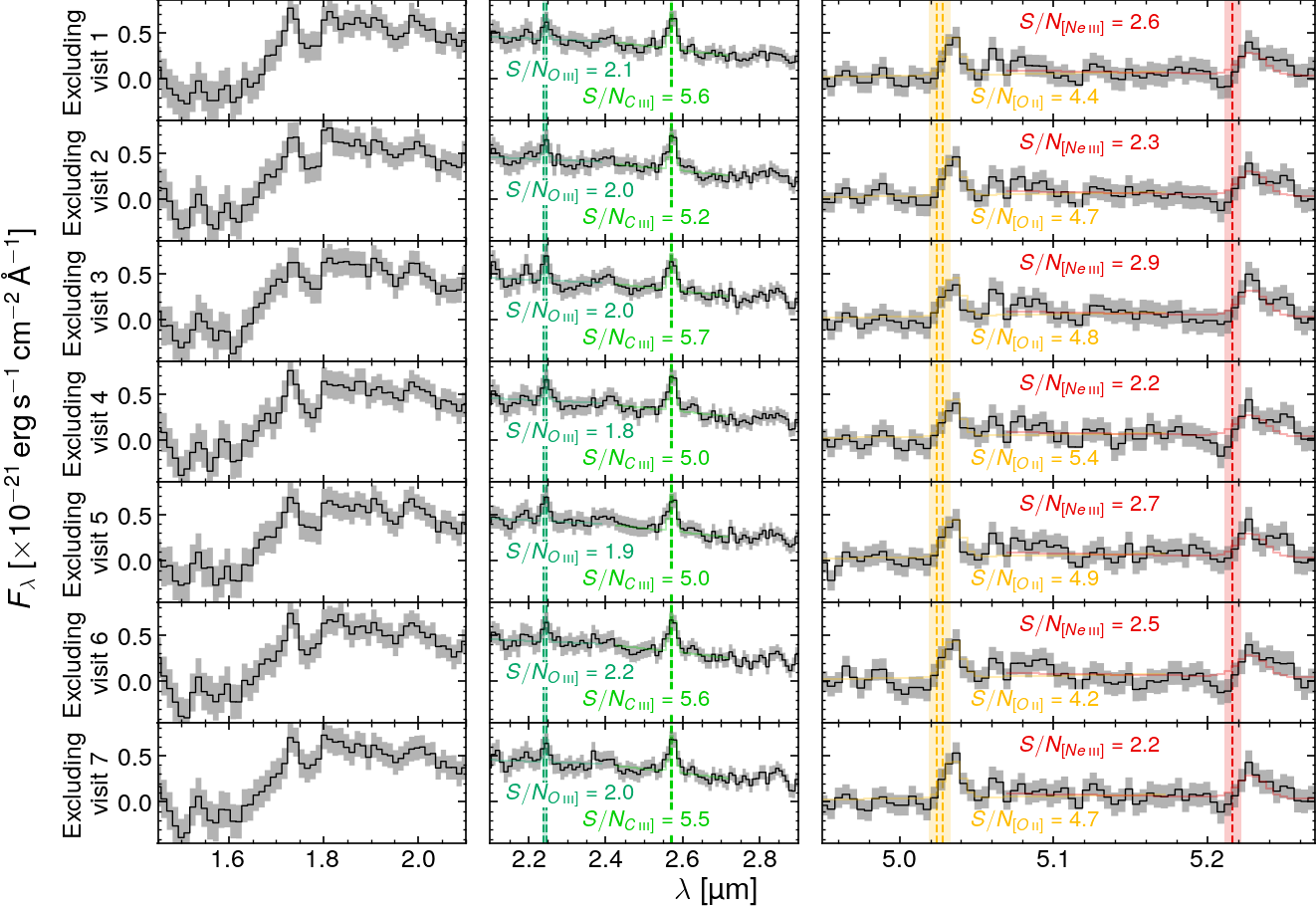}
  \caption{Jackknife combination of the prism data; each row is obtained by
  excluding all the integrations from one of the seven visits and combining the rest.
  The first two rows exclude the first and second visit from PID~1210; the bottom five
  rows each exclude one of the visits from PID~3215. The vertical dashed lines and shaded
  regions are the same as in Figs.~\ref{f.data} and~\ref{f.integs}. The labels report
  the S/N of each line. \CIII appears well detected in all combinations, meaning that
  this line is unlikely to arise from detector artefacts (which may be present in all
  the integrations from the same visit). The different S/N compared to the fiducial fits
  are due to the different model adopted here (see text for more details).
  }
  \label{f.visits}
\end{figure*}

In Fig.~\ref{f.visits} we show the jackknife spectra obtained by stacking
all the individual integrations, but excluding the integrations from a
single visit at a time. This procedure highlights whether the emission lines
we see in the data arise from a detector feature, which may be in common between all the integrations from the same visit.

The stacked spectra are obtained by applying an iterative 3-\textsigma 
clipping algorithm in each wavelength pixel, and by applying inverse-
variance weighting. We fit a single Gaussian with a local linear background
around each of \OIII, \CIII, \OII and \NeIII. Overall, we find that
\CIII is confirmed at 5-\textsigma significance, and does not appear to be
driven by any single visit. The uncertainties about \OIII and \NeIII are
larger than in the fiducial fit, reflecting the increased number of
free parameters used here.

\begin{figure*}
  \includegraphics[width=\textwidth]{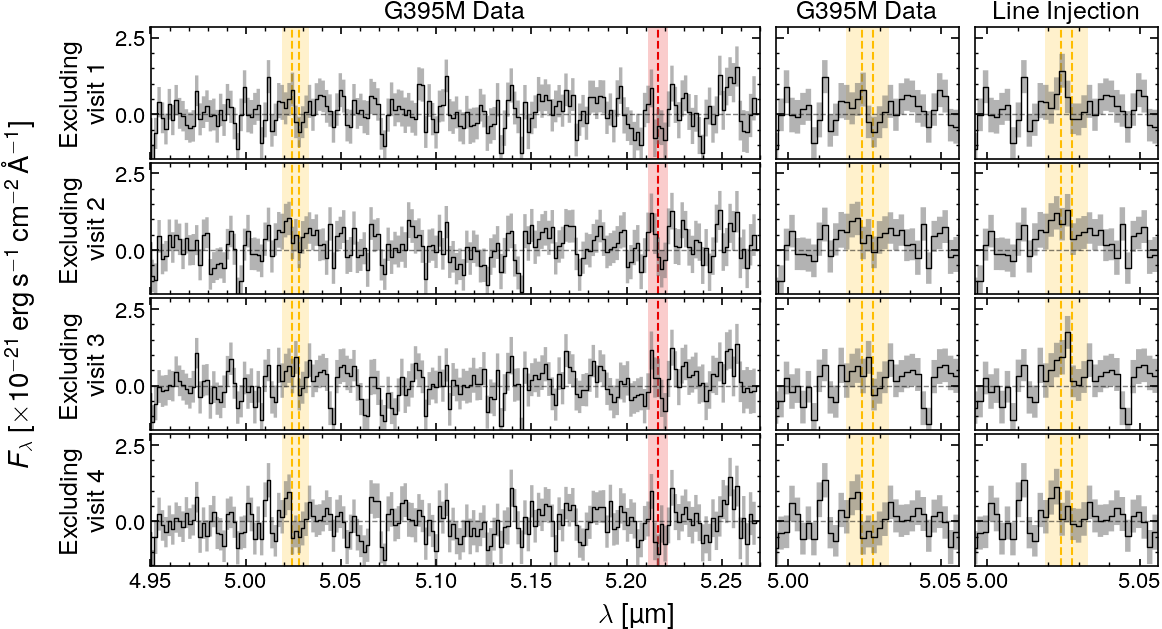}
  \caption{Jackknife combination of the G395M grating data; each row is obtained by
  excluding all the integrations from one of the four visits in PID~3215 and combining
  the rest 
  Table~\ref{t.obs}). The first column displays the spectral region of interest, the
  second column is a zoom-in on the region of \OII, the third column is the same as
  the second, but we added a Gaussian emission line with total flux equal to the \OII flux
  measured in the prism. 
  The vertical dashed lines and shaded
  regions are the same as in Figs.~\ref{f.data} and~\ref{f.integs} and highlight the
  expected wavelengths of \OII (orange) and \NeIII (red). Given the noise of these
  observations, 
  this line is unlikely to arise from detector artefacts (which may be present in all
  the integrations from the same visit). The different S/N compared to the fiducial fits
  are due to the different model adopted here (see text for more details).
  }
  \label{f.visits.grat}
\end{figure*}

We perform the same test in the left column of Fig~\ref{f.visits.grat}, but only for 
PID~3215, because in PID~1210 the G395M exposure time is much shorter than in PID~3215
(only 10~per cent of the total time, Table~\ref{t.obs}).
We find no evidence of strong artefacts in any of the visits. As a recovery test, we add
a single Gaussian emission line with flux equal to the \OII flux measured from the prism,
dispersion equal to one pixel, and centred at
the expected location of the \OII doublet (vertical dashed lines). The second and third
columns in Fig~\ref{f.visits.grat} show a comparison between the data without and with
the artificial line injection. This is the best-case scenario of a single, spectrally 
unresolved emission line; any additional broadening (e.g., due to the doblet nature of
\OII) would make the following estimates more conservative.
We estimate the line sensitivity by integrating the variance spectrum from the pipeline
over two spectral pixels; based on the (conservative) pipeline uncertainties, we should be able to measure this artificial line with S/N=3. This would suggest that at least one of
\OII or \NeIII should be seen in the grating data.
However, the injection test shows that the effective S/N of the artificial line is likely 
lower than 3, and that we would not be able to detect \OII or \NeIII.

\bibliography{hzm_bib}{}
\bibliographystyle{config/aa}

\end{document}